\documentclass[twocolumn, prl, footinbib, floatfix]{revtex4-1}
\usepackage{graphicx,epsfig,array}
\usepackage{amsmath,amssymb,color,subfigure,yllmath,natbib}
\usepackage{newfloat}
\DeclareFloatingEnvironment[name={Fig.}]{suppfigure}
\begin{document}
\title{Non Fermi liquid behavior and continuously tunable resistivity exponents \\
in the Anderson-Hubbard model at finite temperature} 
	\author{Niravkumar D. Patel$^1$}
	\author{Anamitra Mukherjee$^2$}
	\author{Nitin Kaushal$^1$}
	\author{Adriana Moreo$^{1,3}$}
	\author{Elbio Dagotto$^{1,3}$}
	
\affiliation{$^1$Department of Physics and Astronomy, The University of Tennessee, Knoxville, Tennessee 37996, USA}
\affiliation{$^2$School of Physical Sciences, National Institute of Science Education and Research, HBNI, Jatni 752050, India}
\affiliation{$^3$Materials Science and Technology Division, Oak Ridge National Laboratory, Oak Ridge, Tennessee 37831, USA}
\date{\today}
\begin{abstract}
We employ a recently developed computational many-body technique to study for the first time 
the half-filled Anderson-Hubbard model at finite temperature and arbitrary correlation ($U$) 
and disorder ($V$) strengths. Interestingly, the narrow zero temperature 
metallic range induced by disorder from the Mott insulator 
expands with increasing temperature in a manner resembling a quantum critical point. 
Our study of the resistivity temperature scaling $T^{\alpha}$ for this metal 
reveals non Fermi liquid characteristics. Moreover, a continuous dependence  
of $\alpha$ on $U$ and $V$ from linear to nearly quadratic was observed. We argue 
that these exotic results arise from a systematic change with $U$ and $V$ of the 
``effective'' disorder, a combination of quenched disorder and intrinsic localized spins.
\end{abstract}
\maketitle
	
A hallmark of a conventional Fermi liquid (FL) in good metals is the $T^2$ scaling of the resistivity ($\rho$) with temperature $T$. 
However, deviations from this behavior have been reported in several correlated electronic materials 
such as heavy fermions~\cite{PhysRevLett.81.1501,PhysRevLett.85.626,0953-8984-8-48-002,0953-8984-13-35-202}, 
rare earth nickelates~\cite{bad-metal-nickelates}, layered dichalcogenides~\cite{triangle_exp_mft_nandinit_lahoud}, 
and cuprates~\cite{RevModPhys.75.473, Hussey:2003ck,Keimer:bn}. Various ideas for explaining non Fermi liquid (NFL) states 
have  been proposed.  For instance, a $T=0$ quantum critical point (QCP) could induce the linear $\rho \sim T$ scaling in  the cuprates~\cite{Anderson:2006gu,Keimer:bn, PhysRevLett.106.097002}. 
In the NFL observed in the two-dimensional electron gas (2DEG)~\cite{glass-1,glass-2,glass-ps}, 
charge or spin glassy metallic states could provide an alternative~\cite{gt-1,gt-2,gt-3,gtr-1}.  
In spite of these important efforts, the understanding of NFLs in correlated 
systems still eludes theorists. Moreover, in heavy fermion experiments a puzzling continuous 
variation of the $\rho$ vs. $T$ scaling exponent $\alpha$ between 1 and 1.6 
was found~\cite{PhysRevLett.81.1501,PhysRevLett.85.626,0953-8984-8-48-002,0953-8984-13-35-202}. 
Considering that the microscopic physics of the several NFL material families are quite different, it is a challenge 
to find a global understanding of NFL states in correlated systems. In particular, we need to identify
concrete model Hamiltonian systems that not only support NFL states but also, within a single framework,  
capture various NFL systematics observed across different material families.

To address these issues,
here we study the temperature characteristics of the unconventional metal known to 
develop at $T=0$ from the {\it competition} between strong electron interactions and disorder 
in the half-filled Anderson-Hubbard model on a square lattice. In the clean limit,
the ground state is a Mott insulator (MI) and correlated metals 
arising from doping MI's~\cite{PhysRevLett.110.086401} 
violate the $\rho \sim T^2$ scaling. In the other limit where quenched disorder dominates, single particle states are localized in two dimensions and these disorder-induced Anderson insulators often display variable
range hopping behavior  \cite{Imry:2002ww}.

The surprising $T=0$ intermediate metallic state that results 
from the combination of correlations and disorder has been studied theoretically using 
Dynamical Mean Field Theory 
(DMFT)~\cite{Dobrosavljevic:2012fu,stat-dmft_song, stat-dmft-Semmler,r19_screening_dmft_cpa,r20_screening_dmft_cpa}, 
 Quantum Monte Carlo~\cite{r9_dqmc_3d,r22_screening_pqmc,r23_screening_dqmc}, 
Exact Diagonalization~\cite{r24_mit_ED_1d}, and Hartree-Fock~\cite{mft_T0_hed_nan}. 
Experimental results~\cite{PhysRevB.51.7038,r2_mit_exp_dd,r3_mit_exp,cold-atom,triangle_exp_mft_nandinit_lahoud} 
are compatible with the zero temperature calculations. However, the finite temperature 
understanding of this exotic metal and its scaling is limited and several questions
remain. How does a metal 
that arises from competing Mott and Anderson insulators behave at finite temperatures? 
What temperature scaling does the resistivity of the ensuing metal display? 
Is there a dependence of the exponent $\alpha$ on disorder and interaction strengths that
can be tuned? Are spin or charge cluster states~\cite{gtr-1} responsible for such scaling behavior? 
Answers to these open questions are of relevance for experiments and theory alike.

In this publication, we study the half-filled Anderson-Hubbard model at finite 
temperature using the recently developed Mean Field-Monte Carlo (MF-MC) technique. This approach properly incorporates
thermal fluctuations in a mean field theory~\cite{mfmc-cr}. Details and benchmarks are in \textit{Supplementary Material Sec. I}. 
Using MF-MC, here we establish the disorder-interaction-temperature ($V-U-T$) phase diagram. 
In particular, we observed a disorder-induced continuous evolution from the Mott to the Anderson insulators
with a strange metal in between. Our temperature analysis of this region unveils 
an intriguing quasi QCP behavior, with
a narrow metallic region increasing in width with increasing temperature resembling  a 
quantum phase transition.
Through optical conductivity and resistivity calculations, we uncover a striking behavior:  
by changing $U$ and $V$, $\alpha$ can be tuned (akin to heavy fermions) from linear $\alpha =1$,
as in cuprates, to near quadratic. Then disorder
and interactions can be used to modify the scaling $\rho \sim T^{\alpha}$.

\begin{figure}
\centering{
\includegraphics[width=7.5cm, height=7cm, clip=true]{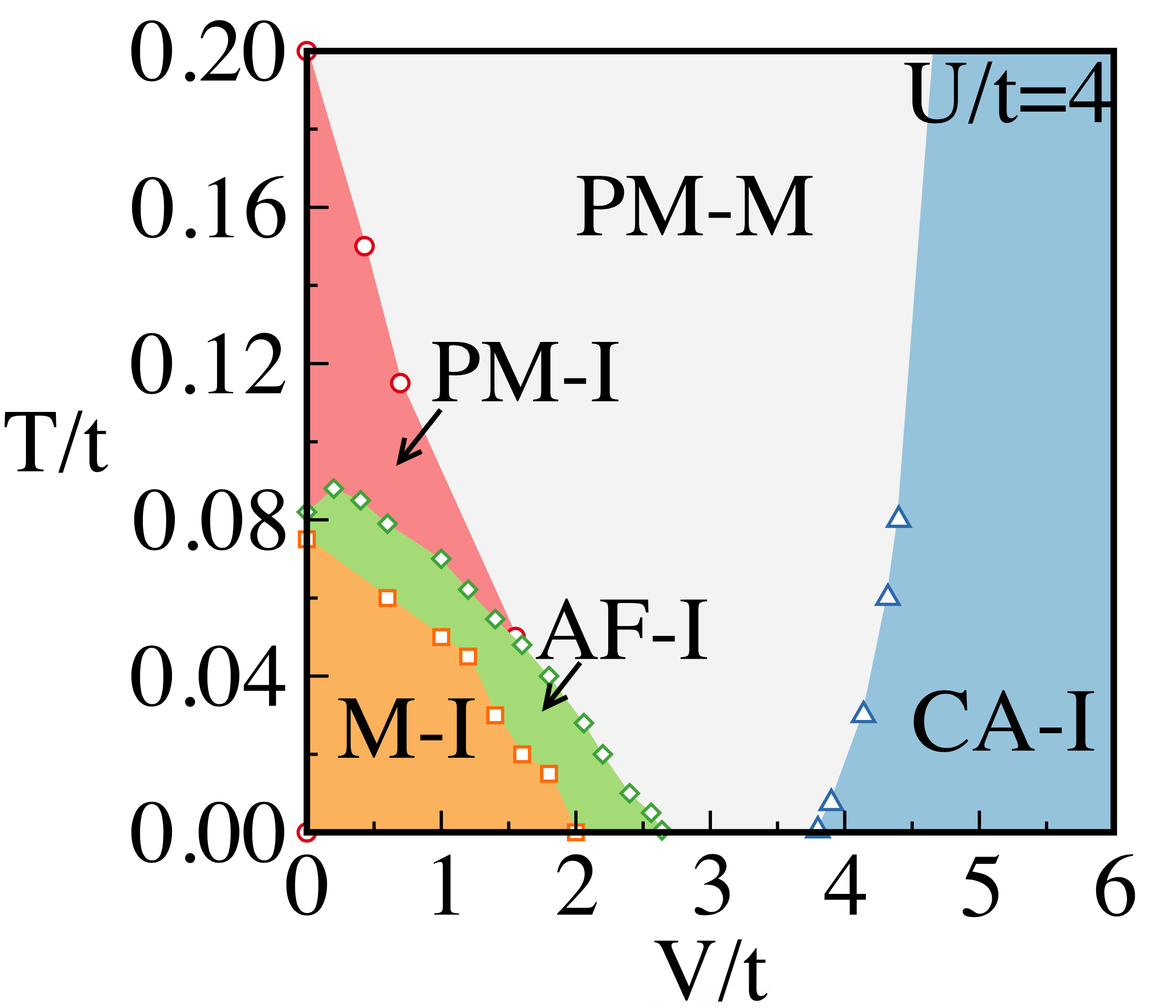}}
\caption{(color online) Temperature ($T$) vs. disorder ($V$) phase diagram at half filling for $U/t=4.0$ obtained 
using a $32^2$ square lattice. The zero temperature antiferromagnetic (AFM) Mott insulator (M-I) to gapless AFM insulator (AF-I) 
at $V\sim 2t$, and the subsequent transition to the paramagnetic metal (PM-M) at $V\sim 3t$, 
are both continuous within our accuracy. For $V>3.8t$ at $T=0$, we observe a disorder-induced correlated Anderson insulator 
(CA-I). Finally for $T/t \in [0.08,0.2]$  and small $V/t$, pink region, we find a gapless PM insulator (PM-I). 
Details are in the text. }
\vspace{-0.0cm}
\label{f-1}
\end{figure}
The model is:
\begin{equation}\nonumber
H = -t \sum_{\langle i, j \rangle \ \sigma} c^{\dagger}_{i,\sigma} c^{\phantom{\dagger}}_{i,\sigma} + \sum_{i} U n_{i \uparrow} n_{i \downarrow} + \sum_{i} (V_{i} - \mu) (n_{i\uparrow} + n_{i\downarrow}),
\label{e1}
\end{equation} 
where the first term is the kinetic energy and the second the standard Hubbard repulsion. 
$c^{\phantom{\dagger}}_{i\sigma}$ ( $c^{\dagger}_{i\sigma}$) annihilates (creates) an electron at site $i$ with spin $\sigma$.
The number operator is $n_{i\sigma} = c^{\dagger}_{i\sigma} c^{\phantom{\dagger}}_{i\sigma}$. The disorder $V_{i}$ at each site is 
chosen randomly in the interval $[-V,V]$ with uniform probability. The chemical potential $\mu$ is 
adjusted to achieve half filling globally. In MF-MC, 
we first Hubbard-Stratonovich decouple the interaction term, by introducing 
a vector auxiliary $\bm{m_i}$ and a scalar $\phi_i$ field at every site. The former couples 
to spin and latter to charge. Dropping the time dependence of the auxiliary fields (Aux. F.) 
a model with ``spin fermion'' characteristics arises. The Aux. F. are treated by classical MC  
that admits thermal fluctuations, and the fermionic sector is solved using Exact Diagonalization.  
Details of the considerable numerical effort involved are discussed in \textit{Supplementary Material Sec I.}

\begin{figure}
	\centering{
		\includegraphics[width=8.5cm, height=8.5cm, clip=true]{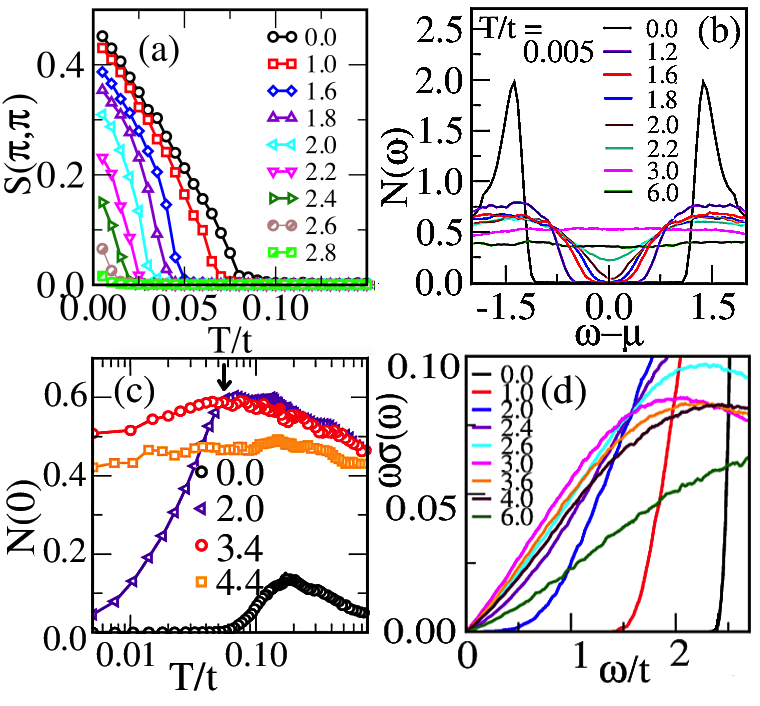}}
	\caption{(color online) Examples of MF-MC data at $U/t=4$ used 
to construct the phase diagram Fig.~\ref{f-1}. (a) Static magnetic structure factor 
at  $\bm{q}=(\pi,\pi)$. Values of $V/t$ are in the column. (b) Low-$T$ density-of-states (DOS)
$N(\omega)$ at various disorder strengths. (c) DOS $\omega=0$ weight at the four $V/t$'s indicated 
vs. $T/t$. Arrow indicates the maximum of $N(0)$ for $V/t=3.4$. (d) $\omega$ times the optical 
conductivity $\sigma(\omega)$, at $T/t=0.005$, for the several $V/t$'s indicated. }
	\vspace{-0.0cm}
	\label{f-2}
\end{figure}

\textit{1. Phase diagram.}  Consider the phase diagram shown in Fig.~\ref{f-1}
at the representative value $U/t=4$. 
Various indicators, such as the $(\pi,\pi)$ static magnetic structure factor, density of states (DOS),  and 
optical conductivity $\sigma(\omega)$ were used (Fig.~\ref{f-2}). At $T/t=0.005$ 
the antiferromagnetic (AFM) order is progressively reduced increasing $V/t$ as shown 
in Fig.~\ref{f-2} (a), and for $V\ge 2.6t$ the signal becomes negligible. At $V=0$, the magnetic order 
starts at $T/t=0.10$ upon cooling but the system remains 
insulating above this temperature, as expected, and a paramagnetic insulator (PM-I), is deduced based on the optical conductivity behavior discussed below. 
Increasing $V/t$, $T_N$ initially slightly 
increases and then reduces with increasing disorder~\cite{upturn}. 
 The metal insulator boundary  decreases roughly linearly with $V/t$, collapsing to zero  at $V/t=2.6$.
Panel (b) shows the low-$T$ DOS, $N(\omega)$, for various disorder strengths. 
We find that the clean-limit Mott gap evolves into a pseudogap gap at $V/t\sim2$. 
This pseudogap persists up to $V/t=2.8$ and flattens out for larger $V/t$, with the weight 
around $\omega=0$ decreasing gradually with disorder as the DOS spreads over a larger energy range 
due to increasing scattering. Thus, the AF-I, PM-M and CA-I phases in Fig.~\ref{f-1} are gapless.

 \begin{figure}
 	\centering{
 		\includegraphics[width=8.5cm, height=7.25cm, clip=true]{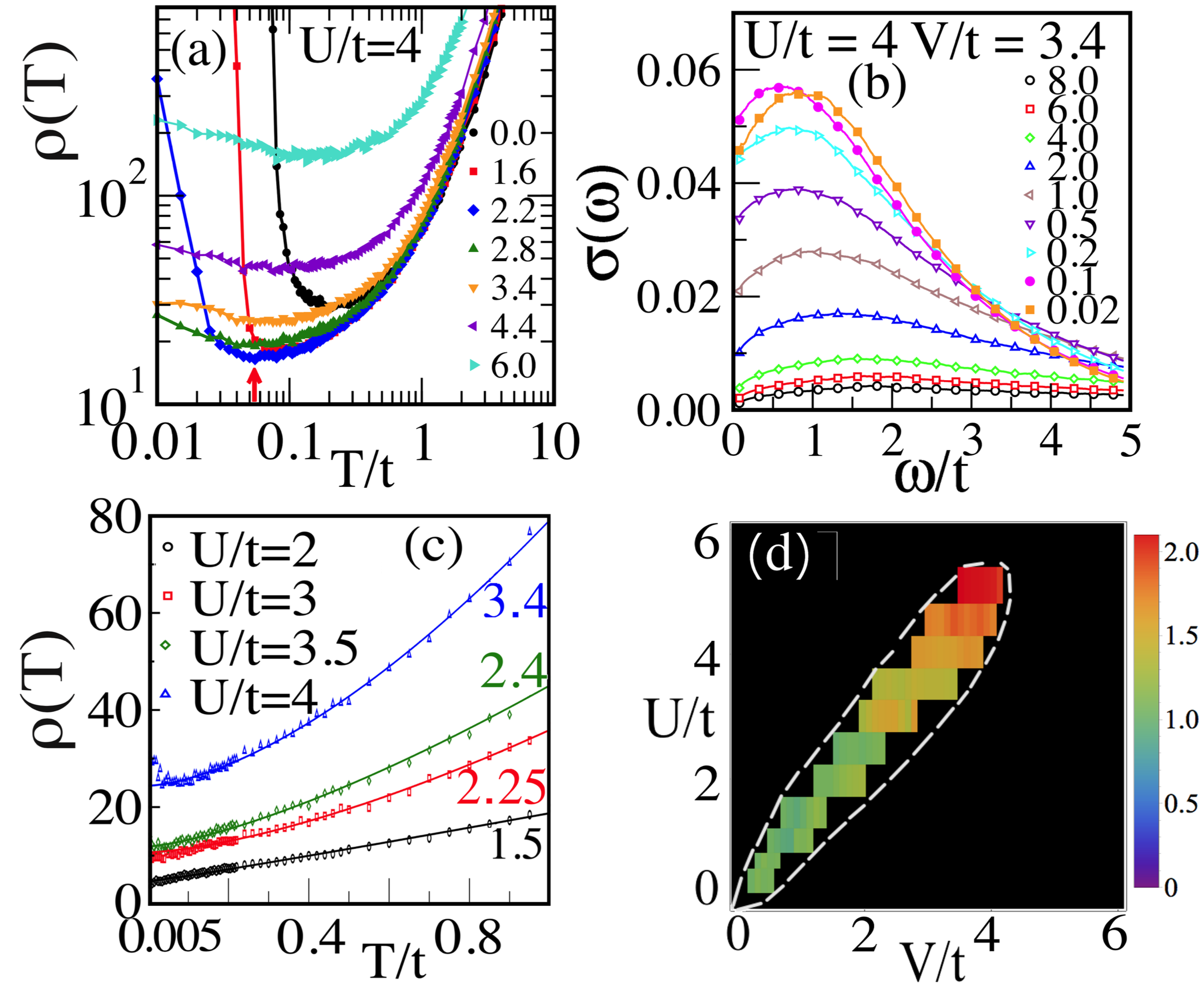}}
 	\caption{(color online)  (a) $\rho$ ($\pi e^2/\hbar a_0$ units) vs. $T/t$
at $U/t=4$ and the various $V/t$'s in the column ($a_0$ is the lattice spacing, set to 1). 
The red arrow at the bottom shows $\rho_{min}$ for $V/t=3.4$.
 		(b) Optical conductivity at $U/t=4$ and $V/t=3.4$ displaying 
non-Drude behavior, at the $T/t$'s in the column. 
(c) $\rho(T)$ at special values of $U/t$ and $V/t$ such that the state is metallic almost in all the
$T/t$ range shown. Solid lines are fits to $\rho(0)+AT^{\alpha}$. For $U/t=2,~3,~3.5$, and $4.0$,
 and the $V/t$ values indicated (color, right), $\alpha$ is  1.02, 1.35, 1.39, and 1.68,
 respectively. (d) The metallic window in the $U-V$ plane at low temperature $T/t=0.005$ 
 is shown by the colored region (the dashed line just guides the eye). The black region denotes insulator. 
 The color of the metallic region depicts the value of $\alpha$ arising from the $T^{\alpha}$ scaling of $\rho(T)$.}
 	\vspace{-0.0cm}
 	\label{f-3}
 \end{figure}
Figure~\ref{f-2} (d) shows $\omega\sigma(\omega)$ at $T=0$  for different disorder values 
covering the Mott-Insulator, the strange metal, and the large disorder (CA-I) phases. 
For $V/t\le2.0$, $\sigma(\omega)$ is clearly gapped. In the narrow range $2.0<V/t<2.6$,  $\omega\sigma(\omega)$ 
tends to zero as $\omega\rightarrow0$ in a nonlinear manner. 
For $2.6\le V/t\le 3.8$,  $\omega\sigma(\omega)\rightarrow 0$ 
linearly i.e. $\sigma(\omega)$ is constant at small $\omega$ indicating a metal. 
For the CA-I,  variable range hopping is expected to provide a $\omega^3ln^3(I/\omega)$ behavior \cite{Imry:2002ww}, 
where $I$ is a typical energy scale depending on the localization length. \textit{Supplementary 
Material Sec II} Fig. 2 shows that the same behavior holds across the finite $T$ insulator to metal transitions as well. Our $T =0$ phase boundaries in Fig. 1 are in excellent agreement with earlier literature~\cite{mft_T0_hed_nan,thesis-dar} and the T = 0 metal is robust against finite size scaling (Supplementary Material, Fig. 3). We  now shift the focus to our main contributions at finite $T$.

\textit{2. Non Fermi liquid metal.} 
The resistivity extracted from $1/\sigma(\omega)$ at small $\omega$ (see supplementary) 
is in Fig.~\ref{f-3} (a) for $U/t=4$. There are several important 
features: (i) $d\rho/dT$ becomes positive at large $T$ for all values of $V/t$; 
(ii) For the PM-I regime at $V/t=0$ and 1.6, $d\rho/dT$ becomes negative  
with eventual divergence at the critical AFM temperature. For the metallic ($V/t=2.8$ and $3.4$) and the CA-I ($V/t=4.4$ 
and $6.6$) phases, $\rho(T)$ saturates at the lowest $T$'s investigated~\cite{finite-note}. (iii) There are resistivity minima 
at finite $T$ which coincide with the corresponding location of the peaks in $N(0)$ in Fig.~\ref{f-2} (c) at, \textit{e.g.}, $V/t=3.4$. 
The NFL nature of the disorder-induced metallic state can be inferred from panel (b) which shows 
that $\sigma(\omega)$ has a non-Drude form with a peak at finite frequency. 
This peak is further pushed to higher frequency with increasing $T$, except at very low $T$ when 
the peak converges to $\omega/t \sim 1$. 

Both $\rho(T)$ here, and specific heat ($C_v$) in \textit{Supplementary Material} Fig. 4, show low $T$ deviations from FL behavior consistent with literature on disorder induced NFL’s~\cite{gtr-1}, justifying our $^\prime$NFL metal$^\prime$  nomenclature.

Consider now the $\rho$ vs. $T$ behavior for the NFL state. In the metallic regime ($V/t=3.4$), 
the resistivity minimum occurs at $T/t\sim 0.1$. From Fig.~\ref{f-2} (c) at $V/t=3.4$, the location 
of $\rho_{min}$ coincides with the peak in $N(0)$ at $T/t\sim 0.055$. This non-monotonic dependence of $N(0)$ on $T$ 
agrees with DQMC studies of the Hubbard model~\cite{paiva} for $V=0$. 
The initial increase of $N(0)$ is due to thermally induced fluctuations that enhance 
the DOS weight at $\omega=0$. At high $T$, the scattering of fermions from  the 
Aux. F.'s suppress $N(0)$, and this non-monotonicity is reflected in the metallic-like 
thermal behavior of $\rho(T)$. In summary, at low $T$ the initial increase of the DOS at the Fermi level 
forces $d\rho/dT$ to be negative, while at high $T$ this DOS is suppressed again because of the localized spins 
and $d\rho/dT$ changes sign.

\textit{3. Scaling of resistivity.} 
In Fig.~\ref{f-3} (c) we show $\rho(T)$ for combinations of $U/t$ and $V/t$ where the system is a metal over a
wide temperature range. 
The full map of the low temperature metallic region in the $U/t-V/t$ plane is in Fig.~\ref{f-3} (d). The resistivity data 
is fitted to $\rho(0)+AT^{\alpha}$ for each case to extract $\alpha$~\cite{fitnote-q}. For small/intermediate 
values of $(U/t,~V/t)$ (open circles), $\rho(T)$ grows linearly with $T$ in the range analyzed. 
For larger $U/t$ 
(and corresponding $V/t$) $\alpha$ increases from $\sim$1.0 to $1.7$ for $U/t=2,~V/t=3.4$. As shown in Fig.~\ref{f-3}  (d), the metallic window at $T/t=0.005$ occurs roughly around the line $U\sim 1.25V$~\cite{RS-comment1}.
%
%
The dashed line guides the eye and 
it  envelops the metallic region. The metallic-regime color scale indicates the value of $\alpha$ 
in the temperature fit of $\rho(T)$. For up to $U/t=2.5$,  $\alpha\sim 1$ growing slowly with $U/t$ (the 
smallest values checked are $U/t=0.5,~V/t=0.5$). For larger $U/t, V/t> 2.5$, $\alpha$ grows reaching a maximum value 
close to 2 for $U/t=5$~\cite{fitnote-p}. This $\alpha\sim2$ does not imply a FL but we believe 
it is just one of the possible transport exponents that occurs in our system in its slow evolution. 
For even larger $U/t$, within our precision the metallic region closes.

\begin{figure}
	\centering{
		\includegraphics[width=8.5cm, height=6.25cm, clip=true]{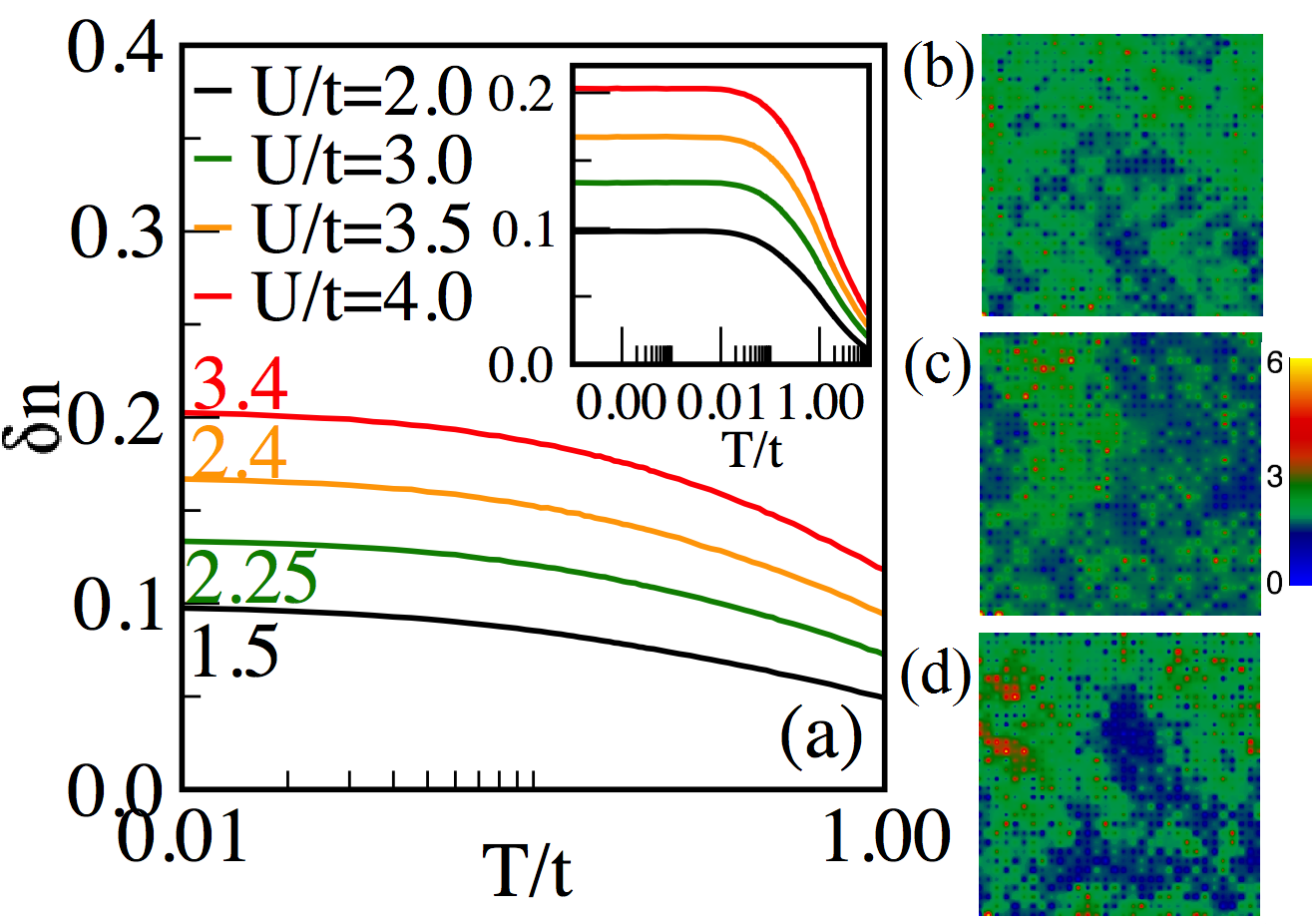}}
	\caption{(color online) Fluctuations in the charge density vs. $T$ at various
$U/t$'s and $V/t$'s (color), in the metallic regime. Inset contains the same data up to $T/t=10$.  
(b,c,d) show real-space maps of $|\psi(\boldmath{r})|^2$ at the Fermi energy using $32^2$ systems 
at $T/t=0.2$ and coupling strengths
($U,~V$) = ($2.0t$, $1.5t$), ($3.0t$, $2.25t$) and ($4.0t$, $3.4t$), respectively. 
Color scale indicates  the range of values for $|\psi(\boldmath{r})|^2$ in $10^{-3}$ units. 
Data in (a) to (d) are averaged over 10 MC samples for a fixed disorder realization.}
	\vspace{-0.0cm}
	\label{f-4}
\end{figure}

\textit{4. Discussion.} 
To better understand the combined effect of disorder and interaction in the metallic phase, 
in Fig.~\ref{f-4} (a) we show the variance of the local density $\{n_i\}$, defined 
as $\delta n(U,~V,~T)=\langle \sqrt{\langle n \rangle^2-\langle n^2 \rangle}\rangle$. The outer angular brackets 
imply averaging over MC samples at fixed $T$, while the inner ones are the quantum average within a single MC sample. 
There are two sources of disorder that control the variance of $\delta n(U,~V,~T)$. First is 
the static disorder ($V$) with variance $\delta V$ and the second are the  Aux. F.'s. These Aux. F.'s directly couple 
to the fermions and indirectly to the disorder through the local fermion occupations (Sec. I of 
the \textit{Supplementary Material}). Then, the Aux. F. also provide an inhomogeneous background that at low temperatures 
follows the presence of intrinsic static disorder. But at temperatures where $k_BT\ge\delta V$, the intrinsic 
disorder ($V$) is unimportant and the Aux. F. configurations become homogeneous in average. 
Thus,  $\delta n(U,~V,~T)$ remains non zero at low $T$ while it tends to zero at large $T$, on MC sample averaging. 
This behavior is observed in the inset of Fig.~\ref{f-4} (a). In the main panel we show the same data 
between $T/t=0.01$ and $1$. We find a systematic increase in $\delta n(U,~V,~T)$ with 
increasing $U/t$ and $V/t$. This variance manifests as real-space charge clusters as shown in Fig.~\ref{f-4} (b) to (d) 
that contains maps of $|\psi(\boldmath{r})|^2$ at fixed $T$, MC sample averaged,  
for states at the Fermi energy.  With increasing $U/t$ and $V/t$, the charge 
clustering and the charge fluctuations magnitude increase systematically following 
the increase in $\delta n(U,~V,~T)$. This provides a controlled enhancement in spatially 
inhomogeneous background from which the fermions scatter~\cite{RS-comment2}.

It is known that fermions coupled to classical variables such as disorder \cite{tca-1}, adiabatic phonons \cite{nfl-phonon} etc. 
can exhibit charge clustering, metallic glasses, and NFL behavior. In our case, not only the quenched disorder but also the 
Aux. F. fluctuations play the role of the classical scatterer that give rise to NFL scaling. 
Such charge clusters and NFL behavior has been experimentally observed in the 2DEG near a $T=0$ quantum critical 
point~\cite{glass-1,glass-2,glass-ps}. Here, within MF-MC, we have found such a ``charge cluster metal'' 
in the half-filled Anderson Hubbard model and also observed that their deviation 
from FL theory can be tuned. This tunability allows us to show that, in a single model Hamiltonian, the 
resistivity scaling with $T$ can vary between linear to near quadratic, features observed in real NFLs 
like cuprates and heavy fermions. 
Our results thus represent progress towards identifying a single model 
system with NFL behavior transcending many material families.   

\textit{Acknowledgments.} A. Mukherjee acknowledges useful discussions with P. Majumdar, 
P. Chakraborty, and H. R. Krishnamurthy.
N.P. and N.K. were supported by the National Science Foundation, under Grant No. DMR-1404375.
E.D. and A. Moreo were supported by the U.S. Department of Energy, Office of Basic Energy Sciences, Materials
Sciences and Engineering Division.

\bibliographystyle{apsrev4-1}
\bibliography{disorder_bib_F} 
\newpage
\clearpage
\begin{center}
	\textbf{Supplementary material}
\end{center}
\section{I.~~ Model and Method}
In this section we discuss the Mean field-Monte Carlo (MF-MC) method. The approach properly incorporates thermal fluctuations in a mean field theory and has been shown to reproduce accurately the properties 
of the clean-limit Hubbard model when compared with Determinantal Quantum Monte Carlo (DQMC) results~\cite{hubb-mcmf}. 
MF-MC allows for efficient simulations on large lattices \cite{ptca} and has also been used to study 
frustrated systems \cite{frus-1,frus-2}, the BCS-BEC crossover \cite{bec-bcs-2,bec-bcs-1}, 
imbalanced Fermi gases \cite{fflo},  and the multiorbital Hubbard model \cite{two-orb-1,2orb-2}. 
Below we present detailed account of the formalism, its numerical implementation, benchmarks against clean system DQMC results, observables calculated and typical parameters (system size, broadening etc.) used in the present calculation.\\

\textit{1. Effective spin-fermion from the Hubbard model: } As mentioned in the main paper, for our study we use the  Anderson-Hubbard model on a square lattice at half filling ($n=1$). The Hamiltonian is:

\begin{equation}
\begin{split}
H &= -t \sum_{\langle i, j \rangle \ \sigma} c^{\dagger}_{i,\sigma} c^{\phantom{\dagger}}_{i,\sigma} + \sum_{i} U n_{i \uparrow} n_{i \downarrow} \\
&+ \sum_{i} (V_{i} - \mu) (n_{i\uparrow} + n_{i\downarrow}),
\end{split}
\label{e1}
\end{equation} 
where the first term is the kinetic energy and the second term is the standard Hubbard repulsion. 
$c^{\phantom{\dagger}}_{i\sigma}$ ( $c^{\dagger}_{i\sigma}$) annihilates (creates) an electron at site $i$ with spin $\sigma$.
The number operator is $n_{i\sigma} = c^{\dagger}_{i\sigma} c^{\phantom{\dagger}}_{i\sigma}$. The disorder $V_{i}$ at 
each site is chosen randomly in the interval $[-V,V]$ with a uniform probability. 
We perform a rotationally invariant decoupling of the interaction term and use the 
Hubbard-Stratonovich (HS) decomposition to obtain an effective Hamiltonian under the 
approximations described below. 
Briefly, the rotational invariant form for the interaction term gives
\begin{eqnarray} 
n_{i,\uparrow}n_{i,\downarrow} &=& \frac{1}{4}(n_i^2)-S_{iz}^2 \nonumber\\
&=& \frac{1}{4}(n_i^2)-({\bold S}_i \cdot \hat{\Omega}_i)^2.
\label{2}
\label{e2}
\end{eqnarray}
Here, the spin operator is ${\bold S_i}=\frac{\hbar}{2}\sum_{\alpha,\beta} c^{\dagger}_{i,\alpha}
{\bold \sigma}^{\phantom\dagger}_{\alpha,\beta}c^{\phantom\dagger}_{i,\beta}$, $\hbar=1$,  
$\{\sigma^x,\sigma^y,\sigma^z \}$ are the Pauli matrices, and $\hat{\Omega}$ is an 
arbitrary unit vector. To decouple this term we introduce two auxiliary 
fields $\bm{m}_i(\tau)$ and $\phi_i(\tau)$ at every site $i$ and every 
imaginary time $\tau$, with each auxiliary field coupled to the spin and the charge degrees of freedom, 
respectively. In our recent work~\cite{hubb-mcmf} we showed that for $V_i=0$, if we consider 
time independent auxiliary fields and replace the $\phi_i$ auxiliary field by its saddle point 
value, $\phi_i=-i\frac{U}{2} \langle n_i\rangle$, we obtain the first three terms of the effective 
Hamiltonian in Eq.~\ref{e3}. Since the disorder term is a one-body interaction 
it remains unchanged from Eq.~\ref{e1} to Eq.~\ref{e3}: 
\begin{equation}
\begin{split}
H_{eff} &= -t \sum_{\langle i, j \rangle \ \sigma} c^{\dagger}_{i,\sigma} c^{\phantom{\dagger}}_{i,\sigma}  
+ \frac{U}{2} \sum_{i} ( \langle n_{i} \rangle n_{i} - \textbf{m}_{i}.\sigma_{i} )  \\
&+ \frac{U}{4} \sum_{i} (\textbf{m}_{i}^{2} - \langle n_{i} \rangle^{2} ) 
+ \sum_{i} (V_{i} - \mu) (n_{i\uparrow} + n_{i\downarrow}).
\end{split}
\label{e3}
\end{equation}

Equation~\ref{e3} is, thus, a one-body Hamiltonian that belongs to the spin fermion class 
as it is a functional of the classical auxiliary fields $\{\bm{m}_i\}$. 
We note that although there is no direct coupling between the auxiliary fields $\{\bm{m}_i\}$ and the disorder, 
there is an indirect coupling through the fermion density which is sensitive to the disorder profile 
and are in turn coupled to the auxiliary fields $\{\bm{m}_i\}$. 

At $T=0$, $H_{eff}$ reduces to a mean field Hamiltonian while at finite $T$, admitting thermal 
fluctuations (via employing a Monte Carlo scheme as discussed below) renders 
the calculation much better than a simple finite $T$ mean-field theory. 
These conclusions are based on results shown below and in previous literature. 
Because of the mixing of mean field theory and Monte Carlo, we call the 
method the ``Mean Field-Monte Carlo'' (MF-MC) approach. 
\\

\textit{2. Treatment of the spin-fermion model: }

Our strategy for solving the resulting spin-fermion model is as follows. We aim to generate 
equilibrium configurations of $\{\bm{m}_i\}$ at a given temperature that minimizes the free energy of the system. 
The free energy of course contains a quantum component for which we diagonalize $H_{eff}$ for a given 
configuration $\{\bm{m}_i\}$. Since we have broken translation invariance by the quenched disorder, 
we cannot set $\langle n_i\rangle=1$ as could be done in the half filling clean problem. Thus, the values 
of $\langle n_i\rangle$ need to be determined at each iteration. Only then we can diagonalize $H_{eff}$. For this task, 
we start with an average value of $\langle n_i\rangle=1$ at the beginning of our calculation at high $T$. 
We then choose a random value set of the Aux.F. and diagonalize $H_{eff}$ and extract the associated 
eigenvalues. We then add the purely classical term $\frac{U}{4} \sum_{i} \textbf{m}_{i}^{2} $ to the 
fermionic energy. The final free energy is then used within a Metropolis algorithm to update 
the $\{\bm{m}_i\}$ variables. By this repeated process the static auxiliary fields $\{\bm{m}_i\}$ are 
annealed with a classical Monte-Carlo (MC). We repeat the classical MC system sweep 50 times. 
We then stop updating the Aux.F. and run a self consistency loop for the $\langle n_i\rangle$ 
in the \textit{fixed} Aux. F. background. Once the $\langle n_i\rangle$ have converged for all 
the sites we stop the self consistency and again run the classical MC for the Aux. F. now for 
the fixed $\langle n_i\rangle$ configuration. The $\langle n_i\rangle$ self consistency is repeated 
after every 50 steps of the classical MC system sweeps during the thermalization portion 
of the classical MC. This usually consists of 2000 MC system sweeps. During the measurement steps, namely
another 2000 MC system sweeps at the same temperature, the density is no longer adjusted, as we found 
that observables are not sensitive to the self consistency after the first 2000~MC system sweeps. 
The process then is repeated at successive lower $T$, where we use the $\langle n_i\rangle$ and Aux. F. 
configurations of the last system sweep of the higher $T$ as the initial condition for the lower temperature.
This is a computational demanding procedure that requires hundreds of nodes because of the fine grid in $U$, $V$,
and $T$ required plus the average over quenched disorder configurations.

To summarize, the static auxiliary fields $\{\bm{m}_i\}$ are annealed with a classical Monte-Carlo (MC) while 
the fermion problem is solved by ED, with some self consistency steps to obtain $\langle n_i\rangle$.   
The chemical potential $\mu$ is freely adjusted to fix the global density to half filling. We performed a total 
of 4000 MC system sweeps at every temperature from which 200 of the last 2000 were used to calculate the observables 
discussed below and in the main paper. We begin MC simulation at high temperature, usually ($T/t = 1.0$) with a random configuration of 
auxiliary variables $\{\bm{m}_{i}\}$ and then cool down to a lowest temperature of ($T/t \sim 5\times10^{-3}$), 
where up to $100$ intermediate temperatures are used to allow for a gradual evolution in the configurations allowing us to 
avoid potential metastable states. The observables are averaged over 10 to 20 quenched disorder realizations. 

To study the effects of random disorder on a strongly interacting system, it is important to consider 
large lattices. We use a variant of the above mentioned ED+MC technique known as the traveling cluster 
approximation (TCA)~\cite{kumar} that can be easily parallelized to access large system sizes~\cite{ptca}. 
We mainly present results for $32^2$ systems although some of our results were cross checked on lattices up to $56^2$.
We used the thermalized configurations of the auxiliary fields and the corresponding eigenvalues and eigenvectors 
to calculate various observables: density of states $N(\omega$), magnetic structure factor in  real and momentum 
space $S(\textbf{q})$ and $C(|{\bold r}|)$, respectively, real-space maps of charge densities ($|\psi(\bm{r_i}, E_F)|^{2}$)) 
for states at the Fermi energy, inverse participation ratio ($|\psi(E_F)|^{4}$) at Fermi energy and optical 
conductivity ($\sigma(\omega)$) using the Kubo-Greenwood formula. We also extract the resistivity ($\rho$) 
vs. temperature. We discuss these observables in detail in the next section.

\begin{suppfigure*}[t]
	\centering{
		\includegraphics[width=17.cm, height=6.cm, clip=true]{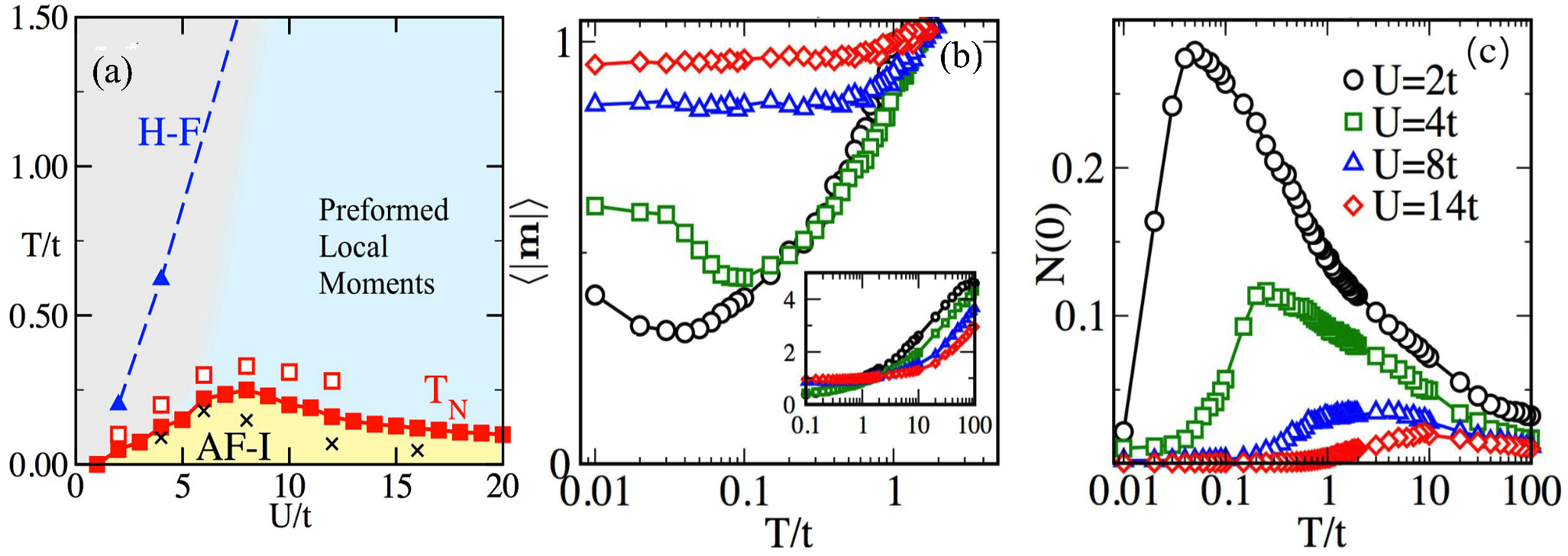}}
	\caption{(color online) (a) The $T/t-U/t$ phase diagram for the one-band Hubbard model. 
		The solid red squares show the dependence of $T_N$ on $U/t$ obtained using the MF-MC technique on $4^3$ clusters. 
		The crosses are estimations of $T_N$ obtained from finite-size scaling. The AF-I region denotes the N\'eel type AFM 
		phase with long-range order and insulating characteristics.  
		The open squares are the $T_N$ obtained from the DQMC method, 
		from  Ref.~\onlinecite{muramatsu-1}.  
		The light blue region depicts the regime of preformed local moments above the AF-I phase. 
		The dashed line shows the $T_N$ obtained from the simplistic Hartree-Fock calculation 
		at finite temperature where the critical temperature incorrectly grows linearly with $U/t$ 
		at large $U/t$. 
		(b) Magnitude of the auxiliary classical fields averaged over the 
		lattice ($\langle|\textbf{m}|\rangle$) vs. temperature for the 
		$U/t$ values shown in (c). At large temperature the thermal fluctuations 
		cause ($\langle|\textbf{m}|\rangle$) to grow linearly with 
		temperature for all the $U$'s shown (inset).  The reason for this 
		temperature dependence of $\langle|\textbf{m}|\rangle$ 
		and its correlation with $N(\omega=0)$ is discussed in the text. 
		(c) Shows $N(\omega=0)$ for 
		different $U/t$ values. This non monotonicity was 
		reported before in a DQMC study, see Ref.~\onlinecite{scalettar-1}. }
	\vspace{-0.0cm}
	\label{sf-1}
\end{suppfigure*}
The MF-MC method can be used to study the Hubbard model at arbitrary coupling and temperature 
and is free of the sign problem. The MF-MC approach has been extensively benchmarked against 
Determinantal Quantum Monte-Carlo for the Hubbard model~\cite{hubb-mcmf} and its $t-t^{\prime}$ variant. 
Also the method can be parallelized to allow access to very large lattices~\cite{ptca} up 
to $250^2$ in two dimensions for the one-band Hubbard model. 

Finally, establishing the phase diagram on a fine $V$, $U$ and $T$ grid, involving also the averaging of the data over 
many disorder realizations, demanded a substantial computational effort. Each disorder configuration, for a single set  of $U$ and $V$ on a 32$^2$ system, requires $\sim 1500$ cpu hours for running the simulation with 100 temperature points. This is multiplied by a factor of 20 for averaging over disorder realizations. Finally, for the full $U$-$V$ phase diagram 400 ($U$, $V$) combinations were used. In addition, calculation of disorder averaged optical conductivity, that scales as $L^3$, $L=32^2$, for every $(U,V)$ set, also added substantial numerical cost. This was possible due to access to the large cluster computing facility (Newton) at the University of Tennessee, Knoxville.

\subsection{Observables}

\textit{(i) Static magnetic structure factor: }Information regarding the N\'eel antiferromagnet 
order is found by computing the static magnetic structure factor. Here we focus on the wavevector $\bm{q}=(\pi,\pi)$ 
as that is the expected order at half-filling for any finite value of $U$. 
The spins  ${\bold S_i}$ are constructed from the eigenvectors of the equilibrated configurations. 
Below the magnetic ordering temperature, the electronic spin degree of freedom are enslaved to magnetic 
auxiliary variables $\bm{m}_i$. Thus, it is interesting to note that even a magnetic structure factor 
constructed only out of the magnetic auxiliary fields closely resembles the full  spin structure factor. The static 
magnetic structure factor is defined as:

\begin{equation}
S({\bold q}) = \frac{1}{N^2} \displaystyle\sum\limits_{i,j} e^{i {\bold q} \cdot ({\bold r}_{i}-{\bold r}_{j} )} 
\langle {{\bold S_i} \cdot {\bold S_j}}\rangle, 
\end{equation} 
where ${\bold q}=\{\pi,\pi\}$ is the wavevector of interest. The normalization $N^2$ is equal to $L^4$, 
where $L$ is the linear dimension of a $L^2$ system.

\textit{(ii) Density of states:} We calculate the density of states (DOS), 
$ N(\omega) = \sum_{m} \delta(\omega-\omega_m)$, 
where $\omega_m$ are the eigenvalues of the fermionic sector and the summation runs up to $2 L^2$, 
i.e. the total number of eigenvalues of a $L^2$ system with spin. $ N(\omega)$ is calculated by implementing
the usual Lorentzian representation of the $\delta$ function. The broadening needed to obtain $N(\omega)$ 
from the Lorentzians is $\sim BW/2L^2$, where $BW$ 
is the fermionic bandwidth at $U=0$. 
Numerically for the 32$^2$ system, the broadening is about $0.004t$.  Hundred $N(\omega)$ samples 
are obtained from the 4000 measurement system sweeps at every temperature. The 200 $N(\omega)$ samples are used to obtain 
the thermally averaged $\langle N(\omega)\rangle_T$ at a fixed temperature. These are further averaged 
over data obtained from 10-20 independent runs with different random number seeds. We also plot $N(\omega-\mu=0)$ vs temperature and disorder.

\textit{(iii) Real space $q=(\pi,\pi)$  magnetic correlations:} We also calculate the real space correlation function between the ${\bold S_i}$ vectors. 
This correlation function is defined as, 
\begin{equation}
C(|{\bold r}|)=\frac{1}{P}\sum\limits_{|{\bold r}|=|{\bold i}-{\bold j}|,a} (-1)^{|{\bold i}-{\bold j}|} \langle S^a_{i}S^a_{j}\rangle. 
\end{equation}
In $C(|{\bold r}|)$  
the summation runs over all P pairs of sites at a distance $|{\bold r}|$
and is normalized accordingly. The sum over $a$ runs over the three directions
$x$, $y$, and $z$. This indicator is used to investigate short range magnetic order in the system in real space.

\textit{(iv)Transport calculations:} The d.c conductivity $\sigma_{dc}$ is estimated by the Kubo-Greenwood
expression \cite{mahan} for the optical conductivity. In a one-electron model
system:
\begin{equation}
\sigma(\omega)=\frac{\pi e^2}{N\hbar a_0}
\sum_{\alpha,\beta} (n_{\alpha} - n_{\beta})
\frac{|f_{\alpha \beta}|^2}{ \epsilon_{\beta} - \epsilon_{\alpha} }
\delta(\omega - (\epsilon_{\beta} - \epsilon_{\alpha})).
\end{equation}
The $f_{\alpha\beta}$ are the matrix elements of the current operator, e.g.,
$\langle \psi_{\alpha} | j_x | \psi_{\beta} \rangle$, and the current operator itself 
(in the tight-binding model) is given by $j_x = i t a_0 e \sum_{i, \sigma} 
(c^{\dagger}_{{i + a_0\hat{x}},\sigma} c^{\phantom{\dagger}}_{i, \sigma} - h.c)$. 
The $\psi_{\alpha}$ are single-particle eigenstates, and $\epsilon_{\alpha}$ are 
the corresponding eigenvalues. The $n_{\alpha}=f(\mu - \epsilon_{\alpha})$ are Fermi factors.

We can compute the low-frequency average, $\sigma_{av}(\mu, \Delta \omega, N) = (\Delta \omega)^{-1} \int_0^{\Delta \omega} 
\sigma(\mu, \omega, N) d\omega$, using periodic boundary conditions in all directions. The averaging interval 
is reduced with increasing $N$, with $\Delta \omega \sim B/N$. Here the constant $B$ is fixed by setting 
$\Delta \omega = 0.008t$ at $N=32^2$. Ideally, the d.c. conductivity is finally obtained 
as $\sigma_{dc}(\mu) = {\lim}_{L \rightarrow \infty} { \sigma}_{av}(\mu, B/L, L)$. However, 
given the extensive numerical cost of our calculation, we simply use the result of $32^2$ system 
as our $\sigma_{dc}(\mu)$. The chemical potential is set to target the required electron density $n$. 
This approach to d.c. transport calculations has been benchmarked in a previous work~\cite{conductivity}. 
\begin{suppfigure*}[t]
	\centering{
		\includegraphics[width=16cm, height=5cm, clip=true]{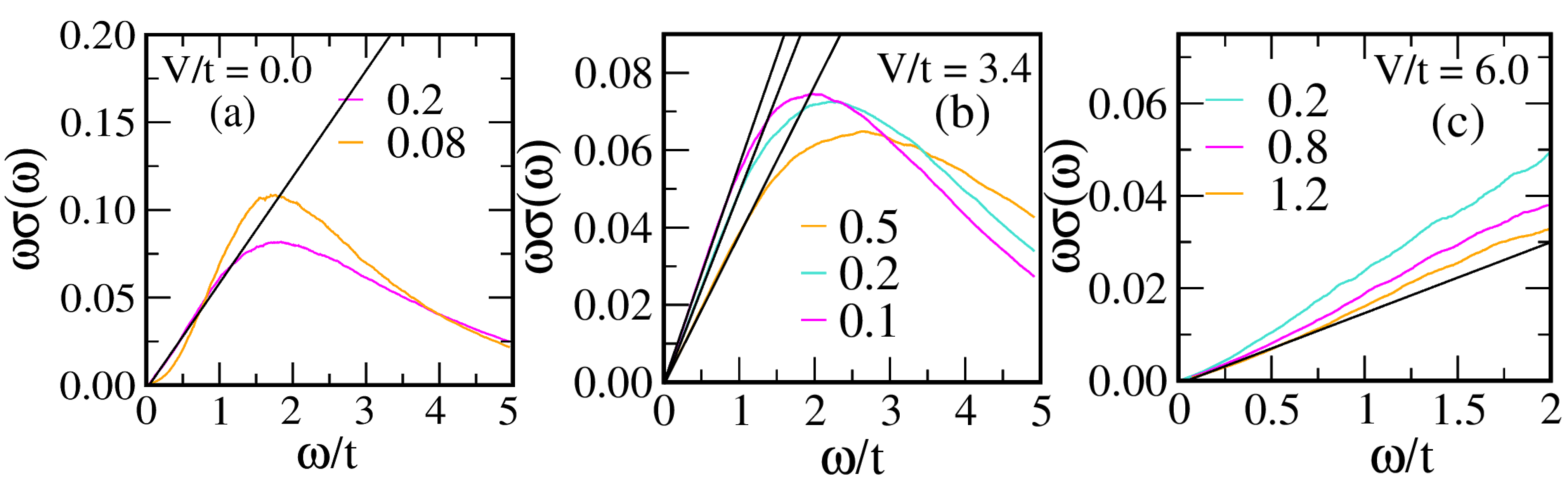}}
	\caption{(color online) Optical conductivity data at finite (high)  temperatures. 
		Plots show $\omega\sigma(\omega)$ vs $\omega$ for $U=4t$, different disorder strengths, 
		and the temperature values indicated for each $V/t$. In the left panel (a) the orange curve displays 
		nonlinear behavior at low $T$ (Mott phase). At higher $T$ we observe a linear behavior, thus a metal. 
		In the middle panel (b), we find linear behavior at all $T$ (i.e. the system is metallic all through). 
		At larger $V$, the high $T$ data shows linearity, indicating a metal, while the low $T$ data show 
		clear deviations from linear behavior. All panels are for $U/t=4$ and using a $32^2$ system. 
		Solid black lines are guides to the eye in all panels.}
	\vspace{-0.0cm}
	\label{sf-2}
\end{suppfigure*}

\subsection{Half-filled Hubbard model: Results $\&$ benchmarking against DQMC}

In this section, we briefly reproduce some pertinent results from earlier publications on the 
finite temperature properties of the half-filled Hubbard model  at zero disorder~\cite{hubb-mcmf}. 
We wish to establish a comparison with DQMC for the finite  $T$ properties and to discuss the nature 
of auxiliary field fluctuations with  $T$. 

\textit{1. Zero disorder phase diagram:} Figure~\ref{sf-1} (a) shows the $U-T$ phase diagram at half filling 
on a three-dimensional $4^3$ lattice from our earlier work~\cite{hubb-mcmf}. The shaded region (below) solid 
square symbols is a $(\pi,\pi,\pi)$ antiferromagnetic Mott insulator. The solid squares are obtained using our 
approach and the open square symbols are DQMC results~\cite{muramatsu-1}. The dashed line with triangles are the $T_N$ 
obtained within finite temperature Hartree-Fock formalism. Clearly, our approach captures the correct non monotonic behavior 
of $T_N$ with $U$ and has semiquantitative agreement with DQMC. \textit{This is much better than simple finite 
	temperature Hartree-Fock that gives qualitatively incorrect results.} In addition the blue shaded region corresponds 
to preformed local moments, another feature that a simplistic finite $T$ mean-field theory cannot capture. 
The gray region to the left of the preformed local moments area is a PM-M. Also as shown in our previous 
work~\cite{hubb-mcmf}, the MF-MC approach can capture the two peak specific heat related to moment formation and moment ordering.

\textit{2. Auxiliary field fluctuations at zero disorder:}  As discussed above,  
the auxiliary fields $\{\bm{m}_i\}$ 
couple linearly to the fermions and they have a quadratic classical cost as well (See Eq. 3). 
In Fig.~\ref{sf-1} (b) we show the system averaged (auxiliary fields) $\{\bm{m}_i\}$ as a function of temperature 
up to $T=100t$ (in the inset) and up to $T=5t$ in the main panel. The data is shown for different $U$'s as indicated 
in panel (b). For $U=4t$ at low $T$, the $\{\bm{m}_i\}$ fields
are enslaved to the magnetic moment and thus have a fixed value. 
With increasing $T$, the magnetic moments weakened and 
so does $\{\bm{m}_i\}$. At $T_N=0.12t$, the system is in a paramagnetic 
metallic state (the gray region in Fig.~\ref{sf-1}). At higher temperatures with negligible 
local moments, the value of  $\langle|\textbf{m}|\rangle$ is governed by thermal fluctuations. At these temperatures 
the auxiliary fields behave as harmonic oscillators with a mean amplitude proportional to $\sqrt{T/U}$. Thus, 
$\langle|\textbf{m}|\rangle$ grows with increasing temperature. With increasing $U$, the dominance of thermal fluctuations 
in governing $\langle|\textbf{m}|\rangle$ is pushed to progressively higher temperatures that increase with $U$.

This unrestricted growth of the Aux. F. at very high $T$ is unavoidable in the continuous Hubbard-Stratonovich 
based MF-MC. However, to compare with DQMC, where the magnitude of the Aux. F are fixed to 1, we restrict the 
comparison to a temperature range for which $\langle|\textbf{m}|\rangle\le 1$. For all data of focus in our publication, particularly
with regards to the temperature scaling of $\rho(T)$, we are in a regime where $\langle|\textbf{m}|\rangle\le 1$.

In panel (c) we show the clean limit $N(0)$ vs $T$. We find that the temperature where $N(0)$ has a peak 
coincides with both the $T_{MIT}$ in Fig.\ref{sf-1} (a) and the temperature where $\langle|\textbf{m}|\rangle$ 
start growing as $\sqrt{(T/U)}$. This suggests that auxiliary field  amplitude fluctuations 
are the main scattering effect in the zero disorder problem at high temperature. This non monotonic dependence 
of $N(0)$ has been found in DQMC calculations as well. In the main text, we show that the same qualitative 
behavior holds even in the presence of disorder. 

\begin{suppfigure}[t]
	\centering{
		\includegraphics[width=6.5cm, height=6cm, clip=true]{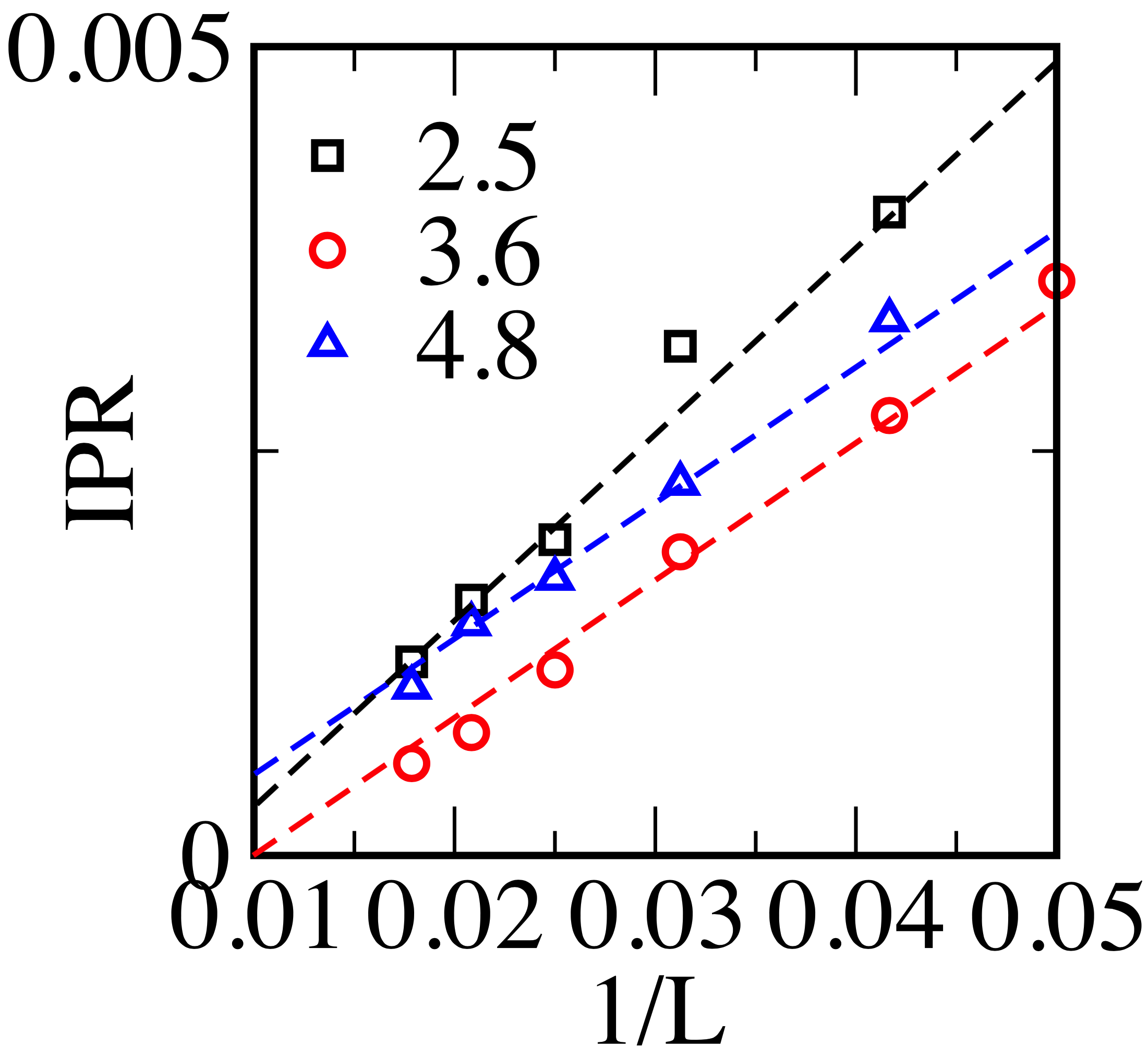}}
	\caption{(color online) Finite size scaling plots for inverse participation ratio for three disorder 
		values $V/t$, at $U/t=4$. $V=2.5$ and 4.8 correspond  to insulating phases while 3.6 corresponds to the metallic phase. 
		The IPR vanishes for the metal and converges to a finite value for the insulator phases as $L\rightarrow\infty$. 
		Dashed lines are a guide to the eye.}
	\vspace{-0.0cm}
	\label{sf-3}
\end{suppfigure}
\section{II.~~Anderson Hubbard model: determining metal insulator boundary}
As mentioned in the main paper, the slope of the resistivity vs temperature is not a good criterion to decide 
between metals and insulators in strongly correlated systems with disorder. Thus, we have used the behavior 
of the optical conductivity vs frequency ($\omega/t$) at small frequency as an indicator of metal insulator transitions. 
As discussed in the main paper, we look for linear behavior of $\omega\sigma(\omega)$ with $\omega$. This ensures 
a constant conductivity close to $\omega/t=0$. Previously at $T=0$ this criterion has been often employed to decide 
between metal and insulator~\cite{mft_T0_hed_nan}. In Fig. 2~(d) of the main paper we have shown the corresponding 
data from our calculations to determine the low temperature (in our case $T/t=0.005$) metal insulator boundary. 
In this Supplementary Material Fig. 2 (a) to (c), we show typical supporting data that we have used to determine 
the finite-temperature metal insulator boundary for $U/t=4$. In (a),  the $V=0$ case is shown. The straight (black) 
line is a guide to the eye. We see a clear departure from linear dependence for $T/t=0.08$, while for $T/t=0.2$ the 
linear behavior is recovered. 
In (b) we show the data for $V/t=3.4$ which is in the metallic regime. Here we see 
that, as in the main text Fig 2 (d), for $V/t=3.4$ at $T/t=0.005$ the  linear behavior continues at all explored 
temperatures (we have checked this up to  $T/t=2$). The highest $T$ data shown here is $T/t=0.5$. In (c) we show that 
the deviation from linearity reduces with increasing $T$ for the CA-I phase as well. In the data shown, $T/t=1.2$ is 
a metal while the lower $T$ system is insulating. 
\begin{suppfigure}[t]
	\centering{
		\includegraphics[width=6.5cm, height=6cm, clip=true]{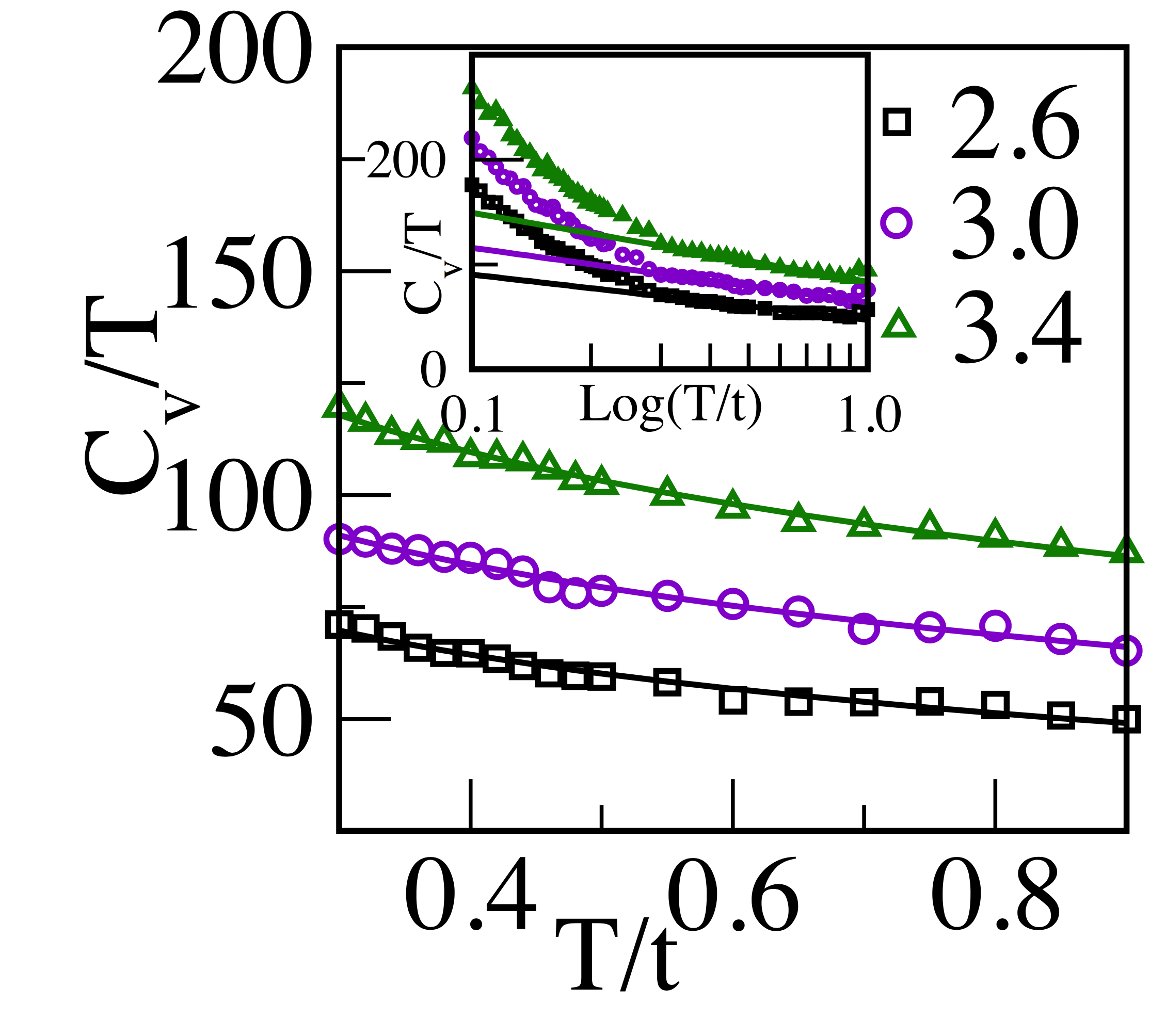}}	
	\caption{(color online) $C_v/T$ vs temperature for the disorder induced metal. The main panel shows 
		data from $T/t=0.3$ to $T/t=0.9$ for three disorder values $V/t$ (indicated) and for $U/t=4$. The solid lines are fits to $\gamma_0Log(T/T_0)$. The inset shows the same data from $T/t=0.1$ to $1.0$. The data for $V/t=2.6$ and $V/t=3.4$ are respectively shifted down and up by 15, in both the main panel and the inset for clarity.}
	\vspace{-0.0cm}
	\label{sf-4}
\end{suppfigure}
At $T/t=0.005$, we also calculated the inverse participation ratio (IPR) for states at the Fermi energy ($E_F$). 
In practice we averaged over a few states in a small window around the $E_F$. This is done for two reasons. Firstly, 
it allows an alternative independent way to cross check the metal insulator boundaries at low $T$. Secondly, 
numerically this calculation is much less expensive than calculating $\sigma(\omega)$. Thus, at least at low $T$ 
we could perform standard finite size scaling to show that IPR extrapolates to  zero for the metallic phase and 
converges to a finite value for both the M-I and the CA-I phases. The data is shown in Fig. 3.

\section{III.~~Specific heat vs temperature}
Figure 4 shows the evolution of the specific heat  ($C_v$) with temperature for the disorder-induced 
metallic phase. As seen in the main panel, $C_v/T$ has a $Log(T/T_0)$ dependence on temperature 
over a wide temperature range.  This behavior holds true for all the three cases of disorder-induced 
metal shown in the figure. The specific heat at $T/t<0.3$ diverges at a faster rate. Thus, our calculation 
also captures another hallmark of a non Fermi liquid, \textit{e.g.} the divergence of the specific heat 
with decreasing temperature, in contrast to FL behavior.

Details of the variations in $\gamma_0$ and $T_0$, in the range of logarithmic 
scaling of $C_v/T$, will be reported elsewhere.

\bibliographystyle{apsrev4-1}
\bibliography{disorder_bib_F} 

\begin{thebibliography}{61}%
\makeatletter
\providecommand \@ifxundefined [1]{%
 \@ifx{#1\undefined}
}%
\providecommand \@ifnum [1]{%
 \ifnum #1\expandafter \@firstoftwo
 \else \expandafter \@secondoftwo
 \fi
}%
\providecommand \@ifx [1]{%
 \ifx #1\expandafter \@firstoftwo
 \else \expandafter \@secondoftwo
 \fi
}%
\providecommand \natexlab [1]{#1}%
\providecommand \enquote  [1]{``#1''}%
\providecommand \bibnamefont  [1]{#1}%
\providecommand \bibfnamefont [1]{#1}%
\providecommand \citenamefont [1]{#1}%
\providecommand \href@noop [0]{\@secondoftwo}%
\providecommand \href [0]{\begingroup \@sanitize@url \@href}%
\providecommand \@href[1]{\@@startlink{#1}\@@href}%
\providecommand \@@href[1]{\endgroup#1\@@endlink}%
\providecommand \@sanitize@url [0]{\catcode `\\12\catcode `\$12\catcode
  `\&12\catcode `\#12\catcode `\^12\catcode `\_12\catcode `\%12\relax}%
\providecommand \@@startlink[1]{}%
\providecommand \@@endlink[0]{}%
\providecommand \url  [0]{\begingroup\@sanitize@url \@url }%
\providecommand \@url [1]{\endgroup\@href {#1}{\urlprefix }}%
\providecommand \urlprefix  [0]{URL }%
\providecommand \Eprint [0]{\href }%
\providecommand \doibase [0]{http://dx.doi.org/}%
\providecommand \selectlanguage [0]{\@gobble}%
\providecommand \bibinfo  [0]{\@secondoftwo}%
\providecommand \bibfield  [0]{\@secondoftwo}%
\providecommand \translation [1]{[#1]}%
\providecommand \BibitemOpen [0]{}%
\providecommand \bibitemStop [0]{}%
\providecommand \bibitemNoStop [0]{.\EOS\space}%
\providecommand \EOS [0]{\spacefactor3000\relax}%
\providecommand \BibitemShut  [1]{\csname bibitem#1\endcsname}%
\let\auto@bib@innerbib\@empty
\bibitem [{\citenamefont {Gegenwart}\ \emph {et~al.}(1998)\citenamefont
  {Gegenwart} \emph {et~al.}}]{PhysRevLett.81.1501}%
  \BibitemOpen
  \bibfield  {author} {\bibinfo {author} {\bibfnamefont {P.}~\bibnamefont
  {Gegenwart}} \emph {et~al.},\ }\href {\doibase 10.1103/PhysRevLett.81.1501}
  {\bibfield  {journal} {\bibinfo  {journal} {Phys. Rev. Lett.}\ }\textbf
  {\bibinfo {volume} {81}},\ \bibinfo {pages} {1501} (\bibinfo {year}
  {1998})}\BibitemShut {NoStop}%
\bibitem [{\citenamefont {Trovarelli}\ \emph {et~al.}(2000)\citenamefont
  {Trovarelli} \emph {et~al.}}]{PhysRevLett.85.626}%
  \BibitemOpen
  \bibfield  {author} {\bibinfo {author} {\bibfnamefont {O.}~\bibnamefont
  {Trovarelli}} \emph {et~al.},\ }\href {\doibase 10.1103/PhysRevLett.85.626}
  {\bibfield  {journal} {\bibinfo  {journal} {Phys. Rev. Lett.}\ }\textbf
  {\bibinfo {volume} {85}},\ \bibinfo {pages} {626} (\bibinfo {year}
  {2000})}\BibitemShut {NoStop}%
\bibitem [{\citenamefont {Julian}\ \emph {et~al.}(1996)\citenamefont {Julian}
  \emph {et~al.}}]{0953-8984-8-48-002}%
  \BibitemOpen
  \bibfield  {author} {\bibinfo {author} {\bibfnamefont {S.~R.}\ \bibnamefont
  {Julian}} \emph {et~al.},\ }\href
  {http://stacks.iop.org/0953-8984/8/i=48/a=002} {\bibfield  {journal}
  {\bibinfo  {journal} {J. Phys.: Condens. Matter}\ }\textbf {\bibinfo {volume}
  {8}},\ \bibinfo {pages} {9675} (\bibinfo {year} {1996})}\BibitemShut
  {NoStop}%
\bibitem [{\citenamefont {Coleman}\ \emph {et~al.}(2001)\citenamefont
  {Coleman}, \citenamefont {Pepin}, \citenamefont {Si},\ and\ \citenamefont
  {Ramazashvili}}]{0953-8984-13-35-202}%
  \BibitemOpen
  \bibfield  {author} {\bibinfo {author} {\bibfnamefont {P.}~\bibnamefont
  {Coleman}}, \bibinfo {author} {\bibfnamefont {C.}~\bibnamefont {Pepin}},
  \bibinfo {author} {\bibfnamefont {Q.}~\bibnamefont {Si}}, \ and\ \bibinfo
  {author} {\bibfnamefont {R.}~\bibnamefont {Ramazashvili}},\ }\href
  {http://stacks.iop.org/0953-8984/13/i=35/a=202} {\bibfield  {journal}
  {\bibinfo  {journal} {J. Phys.: Condens. Matter}\ }\textbf {\bibinfo {volume}
  {13}},\ \bibinfo {pages} {R723} (\bibinfo {year} {2001})}\BibitemShut
  {NoStop}%
\bibitem [{\citenamefont {Jaramillo}\ \emph {et~al.}(2014)\citenamefont
  {Jaramillo}, \citenamefont {Ha}, \citenamefont {Silevitch},\ and\
  \citenamefont {Ramanathan}}]{bad-metal-nickelates}%
  \BibitemOpen
  \bibfield  {author} {\bibinfo {author} {\bibfnamefont {R.}~\bibnamefont
  {Jaramillo}}, \bibinfo {author} {\bibfnamefont {S.~D.}\ \bibnamefont {Ha}},
  \bibinfo {author} {\bibfnamefont {D.~M.}\ \bibnamefont {Silevitch}}, \ and\
  \bibinfo {author} {\bibfnamefont {S.}~\bibnamefont {Ramanathan}},\
  }\href@noop {} {\bibfield  {journal} {\bibinfo  {journal} {Nature Physics}\
  }\textbf {\bibinfo {volume} {10}},\ \bibinfo {pages} {304} (\bibinfo {year}
  {2014})}\BibitemShut {NoStop}%
\bibitem [{\citenamefont {Lahoud}\ \emph {et~al.}(2014)\citenamefont {Lahoud},
  \citenamefont {Meetei}, \citenamefont {Chaska}, \citenamefont {Kanigel},\
  and\ \citenamefont {Trivedi}}]{triangle_exp_mft_nandinit_lahoud}%
  \BibitemOpen
  \bibfield  {author} {\bibinfo {author} {\bibfnamefont {E.}~\bibnamefont
  {Lahoud}}, \bibinfo {author} {\bibfnamefont {O.~N.}\ \bibnamefont {Meetei}},
  \bibinfo {author} {\bibfnamefont {K.~B.}\ \bibnamefont {Chaska}}, \bibinfo
  {author} {\bibfnamefont {A.}~\bibnamefont {Kanigel}}, \ and\ \bibinfo
  {author} {\bibfnamefont {N.}~\bibnamefont {Trivedi}},\ }\href@noop {}
  {\bibfield  {journal} {\bibinfo  {journal} {Phys. Rev. Lett.}\ }\textbf
  {\bibinfo {volume} {112}},\ \bibinfo {pages} {206402} (\bibinfo {year}
  {2014})}\BibitemShut {NoStop}%
\bibitem [{\citenamefont {Damascelli}\ \emph {et~al.}(2003)\citenamefont
  {Damascelli}, \citenamefont {Hussain},\ and\ \citenamefont
  {Shen}}]{RevModPhys.75.473}%
  \BibitemOpen
  \bibfield  {author} {\bibinfo {author} {\bibfnamefont {A.}~\bibnamefont
  {Damascelli}}, \bibinfo {author} {\bibfnamefont {Z.}~\bibnamefont {Hussain}},
  \ and\ \bibinfo {author} {\bibfnamefont {Z.-X.}\ \bibnamefont {Shen}},\
  }\href {\doibase 10.1103/RevModPhys.75.473} {\bibfield  {journal} {\bibinfo
  {journal} {Rev. Mod. Phys.}\ }\textbf {\bibinfo {volume} {75}},\ \bibinfo
  {pages} {473} (\bibinfo {year} {2003})}\BibitemShut {NoStop}%
\bibitem [{\citenamefont {Hussey}\ \emph {et~al.}(2003)\citenamefont {Hussey},
  \citenamefont {Abdel-Jawad}, \citenamefont {Carrington}, \citenamefont
  {Mackenzie},\ and\ \citenamefont {Balicas}}]{Hussey:2003ck}%
  \BibitemOpen
  \bibfield  {author} {\bibinfo {author} {\bibfnamefont {N.~E.}\ \bibnamefont
  {Hussey}}, \bibinfo {author} {\bibfnamefont {M.}~\bibnamefont {Abdel-Jawad}},
  \bibinfo {author} {\bibfnamefont {A.}~\bibnamefont {Carrington}}, \bibinfo
  {author} {\bibfnamefont {A.~P.}\ \bibnamefont {Mackenzie}}, \ and\ \bibinfo
  {author} {\bibfnamefont {L.}~\bibnamefont {Balicas}},\ }\href@noop {}
  {\bibfield  {journal} {\bibinfo  {journal} {Nature}\ }\textbf {\bibinfo
  {volume} {425}},\ \bibinfo {pages} {814} (\bibinfo {year}
  {2003})}\BibitemShut {NoStop}%
\bibitem [{\citenamefont {Keimer}\ \emph {et~al.}(2015)\citenamefont {Keimer},
  \citenamefont {Kivelson}, \citenamefont {Norman}, \citenamefont {Uchida},\
  and\ \citenamefont {Zaanen}}]{Keimer:bn}%
  \BibitemOpen
  \bibfield  {author} {\bibinfo {author} {\bibfnamefont {B.}~\bibnamefont
  {Keimer}}, \bibinfo {author} {\bibfnamefont {S.~A.}\ \bibnamefont
  {Kivelson}}, \bibinfo {author} {\bibfnamefont {M.~R.}\ \bibnamefont
  {Norman}}, \bibinfo {author} {\bibfnamefont {S.}~\bibnamefont {Uchida}}, \
  and\ \bibinfo {author} {\bibfnamefont {J.}~\bibnamefont {Zaanen}},\
  }\href@noop {} {\bibfield  {journal} {\bibinfo  {journal} {Nature}\ }\textbf
  {\bibinfo {volume} {518}},\ \bibinfo {pages} {179} (\bibinfo {year}
  {2015})}\BibitemShut {NoStop}%
\bibitem [{\citenamefont {Anderson}(2006)}]{Anderson:2006gu}%
  \BibitemOpen
  \bibfield  {author} {\bibinfo {author} {\bibfnamefont {P.}~\bibnamefont
  {Anderson}},\ }\href@noop {} {\bibfield  {journal} {\bibinfo  {journal}
  {Nature Physics}\ }\textbf {\bibinfo {volume} {2}},\ \bibinfo {pages} {626}
  (\bibinfo {year} {2006})}\BibitemShut {NoStop}%
\bibitem [{\citenamefont {Casey}\ and\ \citenamefont
  {Anderson}(2011)}]{PhysRevLett.106.097002}%
  \BibitemOpen
  \bibfield  {author} {\bibinfo {author} {\bibfnamefont {P.~A.}\ \bibnamefont
  {Casey}}\ and\ \bibinfo {author} {\bibfnamefont {P.~W.}\ \bibnamefont
  {Anderson}},\ }\href {\doibase 10.1103/PhysRevLett.106.097002} {\bibfield
  {journal} {\bibinfo  {journal} {Phys. Rev. Lett.}\ }\textbf {\bibinfo
  {volume} {106}},\ \bibinfo {pages} {097002} (\bibinfo {year}
  {2011})}\BibitemShut {NoStop}%
\bibitem [{\citenamefont {Bogdanovich}\ and\ \citenamefont {{Popovi\ifmmode
  \acute{c}\else {\'c}\fi{}}}(2002)}]{glass-1}%
  \BibitemOpen
  \bibfield  {author} {\bibinfo {author} {\bibfnamefont {S.~c.~v.}\
  \bibnamefont {Bogdanovich}}\ and\ \bibinfo {author} {\bibfnamefont
  {D.}~\bibnamefont {{Popovi\ifmmode \acute{c}\else {\'c}\fi{}}}},\ }\href
  {\doibase 10.1103/PhysRevLett.88.236401} {\bibfield  {journal} {\bibinfo
  {journal} {Phys. Rev. Lett.}\ }\textbf {\bibinfo {volume} {88}},\ \bibinfo
  {pages} {236401} (\bibinfo {year} {2002})}\BibitemShut {NoStop}%
\bibitem [{\citenamefont {{Jaroszy\ifmmode \acute{n}\else {\'n}\fi{}ski}}\
  \emph {et~al.}(2002)\citenamefont {{Jaroszy\ifmmode \acute{n}\else
  {\'n}\fi{}ski}}, \citenamefont {{Popovi\ifmmode \acute{c}\else {\'c}\fi{}}},\
  and\ \citenamefont {Klapwijk}}]{glass-2}%
  \BibitemOpen
  \bibfield  {author} {\bibinfo {author} {\bibfnamefont {J.}~\bibnamefont
  {{Jaroszy\ifmmode \acute{n}\else {\'n}\fi{}ski}}}, \bibinfo {author}
  {\bibfnamefont {D.}~\bibnamefont {{Popovi\ifmmode \acute{c}\else
  {\'c}\fi{}}}}, \ and\ \bibinfo {author} {\bibfnamefont {T.~M.}\ \bibnamefont
  {Klapwijk}},\ }\href {\doibase 10.1103/PhysRevLett.89.276401} {\bibfield
  {journal} {\bibinfo  {journal} {Phys. Rev. Lett.}\ }\textbf {\bibinfo
  {volume} {89}},\ \bibinfo {pages} {276401} (\bibinfo {year}
  {2002})}\BibitemShut {NoStop}%
\bibitem [{\citenamefont {Kravchenko}\ and\ \citenamefont
  {Sarachik}(2004)}]{glass-ps}%
  \BibitemOpen
  \bibfield  {author} {\bibinfo {author} {\bibfnamefont {S.~V.}\ \bibnamefont
  {Kravchenko}}\ and\ \bibinfo {author} {\bibfnamefont {M.~P.}\ \bibnamefont
  {Sarachik}},\ }\href {http://stacks.iop.org/0034-4885/67/i=1/a=R01}
  {\bibfield  {journal} {\bibinfo  {journal} {Reports on Progress in Physics}\
  }\textbf {\bibinfo {volume} {67}},\ \bibinfo {pages} {1} (\bibinfo {year}
  {2004})}\BibitemShut {NoStop}%
\bibitem [{\citenamefont {Arrachea}\ \emph {et~al.}(2004)\citenamefont
  {Arrachea}, \citenamefont {Dalidovich}, \citenamefont {{Dobrosavljevi\ifmmode
  \acute{c}\else {\'c}\fi{}}},\ and\ \citenamefont {Rozenberg}}]{gt-1}%
  \BibitemOpen
  \bibfield  {author} {\bibinfo {author} {\bibfnamefont {L.}~\bibnamefont
  {Arrachea}}, \bibinfo {author} {\bibfnamefont {D.}~\bibnamefont
  {Dalidovich}}, \bibinfo {author} {\bibfnamefont {V.}~\bibnamefont
  {{Dobrosavljevi\ifmmode \acute{c}\else {\'c}\fi{}}}}, \ and\ \bibinfo
  {author} {\bibfnamefont {M.~J.}\ \bibnamefont {Rozenberg}},\ }\href {\doibase
  10.1103/PhysRevB.69.064419} {\bibfield  {journal} {\bibinfo  {journal} {Phys.
  Rev. B}\ }\textbf {\bibinfo {volume} {69}},\ \bibinfo {pages} {064419}
  (\bibinfo {year} {2004})}\BibitemShut {NoStop}%
\bibitem [{\citenamefont {Dalidovich}\ and\ \citenamefont
  {{Dobrosavljevi\ifmmode \acute{c}\else {\'c}\fi{}}}(2002)}]{gt-2}%
  \BibitemOpen
  \bibfield  {author} {\bibinfo {author} {\bibfnamefont {D.}~\bibnamefont
  {Dalidovich}}\ and\ \bibinfo {author} {\bibfnamefont {V.}~\bibnamefont
  {{Dobrosavljevi\ifmmode \acute{c}\else {\'c}\fi{}}}},\ }\href {\doibase
  10.1103/PhysRevB.66.081107} {\bibfield  {journal} {\bibinfo  {journal} {Phys.
  Rev. B}\ }\textbf {\bibinfo {volume} {66}},\ \bibinfo {pages} {081107}
  (\bibinfo {year} {2002})}\BibitemShut {NoStop}%
\bibitem [{\citenamefont {Sachdev}\ and\ \citenamefont {Read}(1996)}]{gt-3}%
  \BibitemOpen
  \bibfield  {author} {\bibinfo {author} {\bibfnamefont {S.}~\bibnamefont
  {Sachdev}}\ and\ \bibinfo {author} {\bibfnamefont {N.}~\bibnamefont {Read}},\
  }\href {http://stacks.iop.org/0953-8984/8/i=48/a=005} {\bibfield  {journal}
  {\bibinfo  {journal} {J. Phys.: Condens. Matter}\ }\textbf {\bibinfo {volume}
  {8}},\ \bibinfo {pages} {9723} (\bibinfo {year} {1996})}\BibitemShut
  {NoStop}%
\bibitem [{\citenamefont {Miranda}\ and\ \citenamefont
  {Dobrosavljevi{\'c}}(2005)}]{gtr-1}%
  \BibitemOpen
  \bibfield  {author} {\bibinfo {author} {\bibfnamefont {E.}~\bibnamefont
  {Miranda}}\ and\ \bibinfo {author} {\bibfnamefont {V.}~\bibnamefont
  {Dobrosavljevi{\'c}}},\ }\href@noop {} {\bibfield  {journal} {\bibinfo
  {journal} {Reports on Progress in Physics}\ }\textbf {\bibinfo {volume}
  {68}},\ \bibinfo {pages} {2337} (\bibinfo {year} {2005})}\BibitemShut
  {NoStop}%
\bibitem [{\citenamefont {Deng}\ \emph {et~al.}(2013)\citenamefont {Deng},
  \citenamefont {Mravlje}, \citenamefont {{\ifmmode \check{Z}\else
  \v{Z}\fi{}itko}}, \citenamefont {Ferrero}, \citenamefont {Kotliar},\ and\
  \citenamefont {Georges}}]{PhysRevLett.110.086401}%
  \BibitemOpen
  \bibfield  {author} {\bibinfo {author} {\bibfnamefont {X.}~\bibnamefont
  {Deng}}, \bibinfo {author} {\bibfnamefont {J.}~\bibnamefont {Mravlje}},
  \bibinfo {author} {\bibfnamefont {R.}~\bibnamefont {{\ifmmode \check{Z}\else
  \v{Z}\fi{}itko}}}, \bibinfo {author} {\bibfnamefont {M.}~\bibnamefont
  {Ferrero}}, \bibinfo {author} {\bibfnamefont {G.}~\bibnamefont {Kotliar}}, \
  and\ \bibinfo {author} {\bibfnamefont {A.}~\bibnamefont {Georges}},\ }\href
  {\doibase 10.1103/PhysRevLett.110.086401} {\bibfield  {journal} {\bibinfo
  {journal} {Phys. Rev. Lett.}\ }\textbf {\bibinfo {volume} {110}},\ \bibinfo
  {pages} {086401} (\bibinfo {year} {2013})}\BibitemShut {NoStop}%
\bibitem [{\citenamefont {Imry}(2002)}]{Imry:2002ww}%
  \BibitemOpen
  \bibfield  {author} {\bibinfo {author} {\bibfnamefont {Y.}~\bibnamefont
  {Imry}},\ }\href@noop {} {\emph {\bibinfo {title} {{Introduction to
  Mesoscopic Physics}}}}\ (\bibinfo  {publisher} {Oxford University Press on
  Demand},\ \bibinfo {year} {2002})\BibitemShut {NoStop}%
\bibitem [{\citenamefont {Dobrosavljevic}\ \emph {et~al.}(2012)\citenamefont
  {Dobrosavljevic}, \citenamefont {Trivedi},\ and\ \citenamefont {{James M
  Valles}}}]{Dobrosavljevic:2012fu}%
  \BibitemOpen
  \bibfield  {author} {\bibinfo {author} {\bibfnamefont {V.}~\bibnamefont
  {Dobrosavljevic}}, \bibinfo {author} {\bibfnamefont {N.}~\bibnamefont
  {Trivedi}}, \ and\ \bibinfo {author} {\bibfnamefont {J.}~\bibnamefont {{James
  M Valles}}},\ }\href@noop {} {\emph {\bibinfo {title} {{Conductor Insulator
  Quantum Phase Transitions}}}},\ edited by\ \bibinfo {editor} {\bibfnamefont
  {V.}~\bibnamefont {Dobrosavljevic}}, \bibinfo {editor} {\bibfnamefont
  {N.}~\bibnamefont {Trivedi}}, \ and\ \bibinfo {editor} {\bibfnamefont
  {J.~M.}\ \bibnamefont {{Valles Jr}}}\ (\bibinfo  {publisher} {Oxford
  University Press},\ \bibinfo {year} {2012})\BibitemShut {NoStop}%
\bibitem [{\citenamefont {Song}\ \emph {et~al.}(2008)\citenamefont {Song},
  \citenamefont {Wortis},\ and\ \citenamefont {Atkinson}}]{stat-dmft_song}%
  \BibitemOpen
  \bibfield  {author} {\bibinfo {author} {\bibfnamefont {Y.}~\bibnamefont
  {Song}}, \bibinfo {author} {\bibfnamefont {R.}~\bibnamefont {Wortis}}, \ and\
  \bibinfo {author} {\bibfnamefont {W.~A.}\ \bibnamefont {Atkinson}},\
  }\href@noop {} {\bibfield  {journal} {\bibinfo  {journal} {Phys. Rev. B}\
  }\textbf {\bibinfo {volume} {77}},\ \bibinfo {pages} {054202} (\bibinfo
  {year} {2008})}\BibitemShut {NoStop}%
\bibitem [{\citenamefont {Semmler}\ \emph {et~al.}(2010)\citenamefont
  {Semmler}, \citenamefont {Byczuk},\ and\ \citenamefont
  {Hofstetter}}]{stat-dmft-Semmler}%
  \BibitemOpen
  \bibfield  {author} {\bibinfo {author} {\bibfnamefont {D.}~\bibnamefont
  {Semmler}}, \bibinfo {author} {\bibfnamefont {K.}~\bibnamefont {Byczuk}}, \
  and\ \bibinfo {author} {\bibfnamefont {W.}~\bibnamefont {Hofstetter}},\
  }\href@noop {} {\bibfield  {journal} {\bibinfo  {journal} {Phys. Rev. B}\
  }\textbf {\bibinfo {volume} {81}},\ \bibinfo {pages} {115111} (\bibinfo
  {year} {2010})}\BibitemShut {NoStop}%
\bibitem [{\citenamefont {Tanaskovi{\'c}}\ \emph {et~al.}(2003)\citenamefont
  {Tanaskovi{\'c}}, \citenamefont {Dobrosavljevi{\'c}}, \citenamefont
  {Abrahams},\ and\ \citenamefont {Kotliar}}]{r19_screening_dmft_cpa}%
  \BibitemOpen
  \bibfield  {author} {\bibinfo {author} {\bibfnamefont {D.}~\bibnamefont
  {Tanaskovi{\'c}}}, \bibinfo {author} {\bibfnamefont {V.}~\bibnamefont
  {Dobrosavljevi{\'c}}}, \bibinfo {author} {\bibfnamefont {E.}~\bibnamefont
  {Abrahams}}, \ and\ \bibinfo {author} {\bibfnamefont {G.}~\bibnamefont
  {Kotliar}},\ }\href@noop {} {\bibfield  {journal} {\bibinfo  {journal} {Phys.
  Rev. Lett.}\ }\textbf {\bibinfo {volume} {91}},\ \bibinfo {pages} {066603}
  (\bibinfo {year} {2003})}\BibitemShut {NoStop}%
\bibitem [{\citenamefont {Aguiar}\ \emph {et~al.}(2008)\citenamefont {Aguiar},
  \citenamefont {Dobrosavljevi{\'c}}, \citenamefont {Abrahams},\ and\
  \citenamefont {Kotliar}}]{r20_screening_dmft_cpa}%
  \BibitemOpen
  \bibfield  {author} {\bibinfo {author} {\bibfnamefont {M.~C.~O.}\
  \bibnamefont {Aguiar}}, \bibinfo {author} {\bibfnamefont {V.}~\bibnamefont
  {Dobrosavljevi{\'c}}}, \bibinfo {author} {\bibfnamefont {E.}~\bibnamefont
  {Abrahams}}, \ and\ \bibinfo {author} {\bibfnamefont {G.}~\bibnamefont
  {Kotliar}},\ }\href@noop {} {\bibfield  {journal} {\bibinfo  {journal}
  {Physica B: Condensed Matter}\ }\textbf {\bibinfo {volume} {403}},\ \bibinfo
  {pages} {1417} (\bibinfo {year} {2008})}\BibitemShut {NoStop}%
\bibitem [{\citenamefont {Otsuka}\ and\ \citenamefont
  {Hatsugai}(2000)}]{r9_dqmc_3d}%
  \BibitemOpen
  \bibfield  {author} {\bibinfo {author} {\bibfnamefont {Y.}~\bibnamefont
  {Otsuka}}\ and\ \bibinfo {author} {\bibfnamefont {Y.}~\bibnamefont
  {Hatsugai}},\ }\href@noop {} {\bibfield  {journal} {\bibinfo  {journal} {J.
  Phys.: Condens. Matter}\ }\textbf {\bibinfo {volume} {12}},\ \bibinfo {pages}
  {9317} (\bibinfo {year} {2000})}\BibitemShut {NoStop}%
\bibitem [{\citenamefont {Srinivasan}\ \emph {et~al.}(2003)\citenamefont
  {Srinivasan}, \citenamefont {Benenti},\ and\ \citenamefont
  {Shepelyansky}}]{r22_screening_pqmc}%
  \BibitemOpen
  \bibfield  {author} {\bibinfo {author} {\bibfnamefont {B.}~\bibnamefont
  {Srinivasan}}, \bibinfo {author} {\bibfnamefont {G.}~\bibnamefont {Benenti}},
  \ and\ \bibinfo {author} {\bibfnamefont {D.~L.}\ \bibnamefont
  {Shepelyansky}},\ }\href@noop {} {\bibfield  {journal} {\bibinfo  {journal}
  {Phys. Rev. B}\ }\textbf {\bibinfo {volume} {67}},\ \bibinfo {pages} {205112}
  (\bibinfo {year} {2003})}\BibitemShut {NoStop}%
\bibitem [{\citenamefont {Chakraborty}\ \emph {et~al.}(2007)\citenamefont
  {Chakraborty}, \citenamefont {Denteneer},\ and\ \citenamefont
  {Scalettar}}]{r23_screening_dqmc}%
  \BibitemOpen
  \bibfield  {author} {\bibinfo {author} {\bibfnamefont {P.~B.}\ \bibnamefont
  {Chakraborty}}, \bibinfo {author} {\bibfnamefont {P.~J.~H.}\ \bibnamefont
  {Denteneer}}, \ and\ \bibinfo {author} {\bibfnamefont {R.~T.}\ \bibnamefont
  {Scalettar}},\ }\href@noop {} {\bibfield  {journal} {\bibinfo  {journal}
  {Phys. Rev. B}\ }\textbf {\bibinfo {volume} {75}},\ \bibinfo {pages} {125117}
  (\bibinfo {year} {2007})}\BibitemShut {NoStop}%
\bibitem [{\citenamefont {Kotlyar}\ and\ \citenamefont {{Das
  Sarma}}(2001)}]{r24_mit_ED_1d}%
  \BibitemOpen
  \bibfield  {author} {\bibinfo {author} {\bibfnamefont {R.}~\bibnamefont
  {Kotlyar}}\ and\ \bibinfo {author} {\bibfnamefont {S.}~\bibnamefont {{Das
  Sarma}}},\ }\href@noop {} {\bibfield  {journal} {\bibinfo  {journal} {Phys.
  Rev. Lett.}\ }\textbf {\bibinfo {volume} {86}},\ \bibinfo {pages} {2388}
  (\bibinfo {year} {2001})}\BibitemShut {NoStop}%
\bibitem [{\citenamefont {Heidarian}\ and\ \citenamefont
  {Trivedi}(2004)}]{mft_T0_hed_nan}%
  \BibitemOpen
  \bibfield  {author} {\bibinfo {author} {\bibfnamefont {D.}~\bibnamefont
  {Heidarian}}\ and\ \bibinfo {author} {\bibfnamefont {N.}~\bibnamefont
  {Trivedi}},\ }\href@noop {} {\bibfield  {journal} {\bibinfo  {journal} {Phys.
  Rev. Lett.}\ }\textbf {\bibinfo {volume} {93}},\ \bibinfo {pages} {129901}
  (\bibinfo {year} {2004})}\BibitemShut {NoStop}%
\bibitem [{\citenamefont {Kravchenko}\ \emph {et~al.}(1995)\citenamefont
  {Kravchenko}, \citenamefont {Mason}, \citenamefont {Bowker}, \citenamefont
  {Furneaux}, \citenamefont {Pudalov},\ and\ \citenamefont
  {D'Iorio}}]{PhysRevB.51.7038}%
  \BibitemOpen
  \bibfield  {author} {\bibinfo {author} {\bibfnamefont {S.~V.}\ \bibnamefont
  {Kravchenko}}, \bibinfo {author} {\bibfnamefont {W.~E.}\ \bibnamefont
  {Mason}}, \bibinfo {author} {\bibfnamefont {G.~E.}\ \bibnamefont {Bowker}},
  \bibinfo {author} {\bibfnamefont {J.~E.}\ \bibnamefont {Furneaux}}, \bibinfo
  {author} {\bibfnamefont {V.~M.}\ \bibnamefont {Pudalov}}, \ and\ \bibinfo
  {author} {\bibfnamefont {M.}~\bibnamefont {D'Iorio}},\ }\href {\doibase
  10.1103/PhysRevB.51.7038} {\bibfield  {journal} {\bibinfo  {journal} {Phys.
  Rev. B}\ }\textbf {\bibinfo {volume} {51}},\ \bibinfo {pages} {7038}
  (\bibinfo {year} {1995})}\BibitemShut {NoStop}%
\bibitem [{\citenamefont {Sarma}\ \emph {et~al.}(1998)\citenamefont {Sarma}
  \emph {et~al.}}]{r2_mit_exp_dd}%
  \BibitemOpen
  \bibfield  {author} {\bibinfo {author} {\bibfnamefont {D.~D.}\ \bibnamefont
  {Sarma}} \emph {et~al.},\ }\href@noop {} {\bibfield  {journal} {\bibinfo
  {journal} {Phys. Rev. Lett.}\ }\textbf {\bibinfo {volume} {80}},\ \bibinfo
  {pages} {4004} (\bibinfo {year} {1998})}\BibitemShut {NoStop}%
\bibitem [{\citenamefont {Nakatsuji}\ \emph {et~al.}(2004)\citenamefont
  {Nakatsuji}, \citenamefont {Dobrosavljevi{\'c}}, \citenamefont
  {Tanaskovi{\'c}}, \citenamefont {Minakata}, \citenamefont {Fukazawa},\ and\
  \citenamefont {Maeno}}]{r3_mit_exp}%
  \BibitemOpen
  \bibfield  {author} {\bibinfo {author} {\bibfnamefont {S.}~\bibnamefont
  {Nakatsuji}}, \bibinfo {author} {\bibfnamefont {V.}~\bibnamefont
  {Dobrosavljevi{\'c}}}, \bibinfo {author} {\bibfnamefont {D.}~\bibnamefont
  {Tanaskovi{\'c}}}, \bibinfo {author} {\bibfnamefont {M.}~\bibnamefont
  {Minakata}}, \bibinfo {author} {\bibfnamefont {H.}~\bibnamefont {Fukazawa}},
  \ and\ \bibinfo {author} {\bibfnamefont {Y.}~\bibnamefont {Maeno}},\
  }\href@noop {} {\bibfield  {journal} {\bibinfo  {journal} {Phys. Rev. Lett.}\
  }\textbf {\bibinfo {volume} {93}},\ \bibinfo {pages} {146401} (\bibinfo
  {year} {2004})}\BibitemShut {NoStop}%
\bibitem [{\citenamefont {Sanchez-Palencia}\ and\ \citenamefont
  {Lewenstein}(2010)}]{cold-atom}%
  \BibitemOpen
  \bibfield  {author} {\bibinfo {author} {\bibfnamefont {L.}~\bibnamefont
  {Sanchez-Palencia}}\ and\ \bibinfo {author} {\bibfnamefont {M.}~\bibnamefont
  {Lewenstein}},\ }\href@noop {} {\bibfield  {journal} {\bibinfo  {journal}
  {Nature Physics}\ }\textbf {\bibinfo {volume} {6}},\ \bibinfo {pages} {87}
  (\bibinfo {year} {2010})}\BibitemShut {NoStop}%
\bibitem [{mfm()}]{mfmc-cr}%
  \BibitemOpen
  \href@noop {} {}\bibinfo {note} {While our many-body techniques capture
  order-parameter fluctuations, they do not include spin-flip processes
  characteristics of Kondo screening effects of local moments, \textit{e.g.} E.
  Miranda \textit{et al.}, Phys. Rev. Lett. \textbf{78}, 290 (1997) and S. Sen
  \textit{et al.}, Phys. Rev. B \textbf{94}, 235104 (2016). Thus our effort is
  complementary to other previous calculations.}\BibitemShut {Stop}%
\bibitem [{upt()}]{upturn}%
  \BibitemOpen
  \href@noop {} {}\bibinfo {note} {The slight upturn in $T_N$ at small $V/t$ is
  compatible with the discussion in M. Ulmke, \textit{et. al.} Phys. Rev. B
  \textbf{51} 10411 (1995).}\BibitemShut {Stop}%
\bibitem [{\citenamefont {Heidarian}(2006)}]{thesis-dar}%
  \BibitemOpen
  \bibfield  {author} {\bibinfo {author} {\bibfnamefont {D.}~\bibnamefont
  {Heidarian}},\ }\emph {\bibinfo {title} {{Metal-Insulator Transitions In Two
  Dimensions}}},\ \href@noop {} {Ph.D. thesis},\ \bibinfo  {school} {School of
  Natural Sciences Tata Institute of Fundamental Research Mumbai} (\bibinfo
  {year} {2006})\BibitemShut {NoStop}%
\bibitem [{fin()}]{finite-note}%
  \BibitemOpen
  \href@noop {} {}\bibinfo {note} {Since our calculation is on finite system
  and the lowest temperature accessed is $T=0.005t$, the expected divergence in
  $\rho(T)$ as $T\rightarrow0$ for the CA-I is not captured. Thus the optical
  conductivity data is the best indicator of metal vs insulator.}\BibitemShut
  {Stop}%
\bibitem [{\citenamefont {Paiva}\ \emph {et~al.}(2011)\citenamefont {Paiva},
  \citenamefont {Loh}, \citenamefont {Randeria}, \citenamefont {Scalettar},\
  and\ \citenamefont {Trivedi}}]{paiva}%
  \BibitemOpen
  \bibfield  {author} {\bibinfo {author} {\bibfnamefont {T.}~\bibnamefont
  {Paiva}}, \bibinfo {author} {\bibfnamefont {Y.~L.}\ \bibnamefont {Loh}},
  \bibinfo {author} {\bibfnamefont {M.}~\bibnamefont {Randeria}}, \bibinfo
  {author} {\bibfnamefont {R.~T.}\ \bibnamefont {Scalettar}}, \ and\ \bibinfo
  {author} {\bibfnamefont {N.}~\bibnamefont {Trivedi}},\ }\href {\doibase
  10.1103/PhysRevLett.107.086401} {\bibfield  {journal} {\bibinfo  {journal}
  {Phys. Rev. Lett.}\ }\textbf {\bibinfo {volume} {107}},\ \bibinfo {pages}
  {086401} (\bibinfo {year} {2011})}\BibitemShut {NoStop}%
\bibitem [{fit({\natexlab{a}})}]{fitnote-q}%
  \BibitemOpen
  \href@noop {} {} \bibinfo {note} {There is a also a
  variation in $\alpha$ with $V$ for a fixed $U$, details will be reported
  elsewhere as this is not the central issue being addressed here.}\BibitemShut
  {Stop}%
\bibitem [{RS-({\natexlab{a}})}]{RS-comment1}%
  \BibitemOpen
  \href@noop {} {}  \bibinfo {note} {In the absence of
  quantum ($t=0$) and thermal ($T=0$) fluctuations, the line $U=V$ is special,
  separating a phase with {\it all} sites singly occupied from another phase
  with a mixture of occupancies. This is why our metallic phase is close to
  $U=V$.}\BibitemShut {Stop}%
\bibitem [{fit({\natexlab{b}})}]{fitnote-p}%
  \BibitemOpen
  \href@noop {} {} \bibinfo {note} {Resistivity minimum
  occurs for all the cases but is pushed to lower temperatures for smaller $U$,
  so the fitting is done in the regime where $d\rho/dT>0$ going up to
  $T/t=1$.}\BibitemShut {Stop}%
\bibitem [{RS-({\natexlab{b}})}]{RS-comment2}%
  \BibitemOpen
  \href@noop {} {} \bibinfo {note} {Extensions of the
  present work may include other forms of disorder (hoppings, etc.) as explored
  in literature \cite{scale-s1,scale-s2}}\BibitemShut {NoStop}%
\bibitem [{\citenamefont {Kumar}\ and\ \citenamefont
  {Majumdar}(2005{\natexlab{a}})}]{tca-1}%
  \BibitemOpen
  \bibfield  {author} {\bibinfo {author} {\bibfnamefont {S.}~\bibnamefont
  {Kumar}}\ and\ \bibinfo {author} {\bibfnamefont {P.}~\bibnamefont
  {Majumdar}},\ }\href {\doibase 10.1103/PhysRevLett.94.136601} {\bibfield
  {journal} {\bibinfo  {journal} {Phys. Rev. Lett.}\ }\textbf {\bibinfo
  {volume} {94}},\ \bibinfo {pages} {136601} (\bibinfo {year}
  {2005}{\natexlab{a}})}\BibitemShut {NoStop}%
\bibitem [{\citenamefont {Blawid}\ \emph {et~al.}(2003)\citenamefont {Blawid},
  \citenamefont {Deppeler},\ and\ \citenamefont {Millis}}]{nfl-phonon}%
  \BibitemOpen
  \bibfield  {author} {\bibinfo {author} {\bibfnamefont {S.}~\bibnamefont
  {Blawid}}, \bibinfo {author} {\bibfnamefont {A.}~\bibnamefont {Deppeler}}, \
  and\ \bibinfo {author} {\bibfnamefont {A.~J.}\ \bibnamefont {Millis}},\
  }\href {\doibase 10.1103/PhysRevB.67.165105} {\bibfield  {journal} {\bibinfo
  {journal} {Phys. Rev. B}\ }\textbf {\bibinfo {volume} {67}},\ \bibinfo
  {pages} {165105} (\bibinfo {year} {2003})}\BibitemShut {NoStop}%
\bibitem [{\citenamefont {Denteneer}\ \emph {et~al.}(2001)\citenamefont
  {Denteneer}, \citenamefont {Scalettar},\ and\ \citenamefont
  {Trivedi}}]{scale-s1}%
  \BibitemOpen
  \bibfield  {author} {\bibinfo {author} {\bibfnamefont {P.~J.~H.}\
  \bibnamefont {Denteneer}}, \bibinfo {author} {\bibfnamefont {R.~T.}\
  \bibnamefont {Scalettar}}, \ and\ \bibinfo {author} {\bibfnamefont
  {N.}~\bibnamefont {Trivedi}},\ }\href {\doibase
  10.1103/PhysRevLett.87.146401} {\bibfield  {journal} {\bibinfo  {journal}
  {Phys. Rev. Lett.}\ }\textbf {\bibinfo {volume} {87}},\ \bibinfo {pages}
  {146401} (\bibinfo {year} {2001})}\BibitemShut {NoStop}%
\bibitem [{\citenamefont {Denteneer}\ and\ \citenamefont
  {Scalettar}(2003)}]{scale-s2}%
  \BibitemOpen
  \bibfield  {author} {\bibinfo {author} {\bibfnamefont {P.~J.~H.}\
  \bibnamefont {Denteneer}}\ and\ \bibinfo {author} {\bibfnamefont {R.~T.}\
  \bibnamefont {Scalettar}},\ }\href {\doibase 10.1103/PhysRevLett.90.246401}
  {\bibfield  {journal} {\bibinfo  {journal} {Phys. Rev. Lett.}\ }\textbf
  {\bibinfo {volume} {90}},\ \bibinfo {pages} {246401} (\bibinfo {year}
  {2003})}\BibitemShut {NoStop}%
\bibitem [{\citenamefont {Mukherjee}\ \emph {et~al.}(2014)\citenamefont
  {Mukherjee}, \citenamefont {Patel}, \citenamefont {Dong}, \citenamefont
  {Johnston}, \citenamefont {Moreo},\ and\ \citenamefont
  {Dagotto}}]{hubb-mcmf}%
  \BibitemOpen
  \bibfield  {author} {\bibinfo {author} {\bibfnamefont {A.}~\bibnamefont
  {Mukherjee}}, \bibinfo {author} {\bibfnamefont {N.~D.}\ \bibnamefont
  {Patel}}, \bibinfo {author} {\bibfnamefont {S.}~\bibnamefont {Dong}},
  \bibinfo {author} {\bibfnamefont {S.}~\bibnamefont {Johnston}}, \bibinfo
  {author} {\bibfnamefont {A.}~\bibnamefont {Moreo}}, \ and\ \bibinfo {author}
  {\bibfnamefont {E.}~\bibnamefont {Dagotto}},\ }\href {\doibase
  10.1103/PhysRevB.90.205133} {\bibfield  {journal} {\bibinfo  {journal} {Phys.
  Rev. B}\ }\textbf {\bibinfo {volume} {90}},\ \bibinfo {pages} {205133}
  (\bibinfo {year} {2014})}\BibitemShut {NoStop}%
\bibitem [{\citenamefont {Mukherjee}\ \emph {et~al.}(2015)\citenamefont
  {Mukherjee}, \citenamefont {Patel}, \citenamefont {Bishop},\ and\
  \citenamefont {Dagotto}}]{ptca}%
  \BibitemOpen
  \bibfield  {author} {\bibinfo {author} {\bibfnamefont {A.}~\bibnamefont
  {Mukherjee}}, \bibinfo {author} {\bibfnamefont {N.~D.}\ \bibnamefont
  {Patel}}, \bibinfo {author} {\bibfnamefont {C.}~\bibnamefont {Bishop}}, \
  and\ \bibinfo {author} {\bibfnamefont {E.}~\bibnamefont {Dagotto}},\ }\href
  {\doibase 10.1103/PhysRevE.91.063303} {\bibfield  {journal} {\bibinfo
  {journal} {Phys. Rev. E}\ }\textbf {\bibinfo {volume} {91}},\ \bibinfo
  {pages} {063303} (\bibinfo {year} {2015})}\BibitemShut {NoStop}%
\bibitem [{\citenamefont {Tiwari}\ and\ \citenamefont
  {Majumdar}(2014)}]{frus-1}%
  \BibitemOpen
  \bibfield  {author} {\bibinfo {author} {\bibfnamefont {R.}~\bibnamefont
  {Tiwari}}\ and\ \bibinfo {author} {\bibfnamefont {P.}~\bibnamefont
  {Majumdar}},\ }\href {http://stacks.iop.org/0295-5075/108/i=2/a=27007}
  {\bibfield  {journal} {\bibinfo  {journal} {EPL (Europhysics Letters)}\
  }\textbf {\bibinfo {volume} {108}},\ \bibinfo {pages} {27007} (\bibinfo
  {year} {2014})}\BibitemShut {NoStop}%
\bibitem [{\citenamefont {Swain}\ \emph {et~al.}(2016)\citenamefont {Swain},
  \citenamefont {Tiwari},\ and\ \citenamefont {Majumdar}}]{frus-2}%
  \BibitemOpen
  \bibfield  {author} {\bibinfo {author} {\bibfnamefont {N.}~\bibnamefont
  {Swain}}, \bibinfo {author} {\bibfnamefont {R.}~\bibnamefont {Tiwari}}, \
  and\ \bibinfo {author} {\bibfnamefont {P.}~\bibnamefont {Majumdar}},\ }\href
  {\doibase 10.1103/PhysRevB.94.155119} {\bibfield  {journal} {\bibinfo
  {journal} {Phys. Rev. B}\ }\textbf {\bibinfo {volume} {94}},\ \bibinfo
  {pages} {155119} (\bibinfo {year} {2016})}\BibitemShut {NoStop}%
\bibitem [{\citenamefont {Karmakar}\ and\ \citenamefont
  {Majumdar}(2016{\natexlab{a}})}]{bec-bcs-2}%
  \BibitemOpen
  \bibfield  {author} {\bibinfo {author} {\bibfnamefont {M.}~\bibnamefont
  {Karmakar}}\ and\ \bibinfo {author} {\bibfnamefont {P.}~\bibnamefont
  {Majumdar}},\ }\href {\doibase 10.1103/PhysRevA.93.053609} {\bibfield
  {journal} {\bibinfo  {journal} {Phys. Rev. A}\ }\textbf {\bibinfo {volume}
  {93}},\ \bibinfo {pages} {053609} (\bibinfo {year}
  {2016}{\natexlab{a}})}\BibitemShut {NoStop}%
\bibitem [{\citenamefont {Tarat}\ and\ \citenamefont
  {Majumdar}(2015)}]{bec-bcs-1}%
  \BibitemOpen
  \bibfield  {author} {\bibinfo {author} {\bibfnamefont {S.}~\bibnamefont
  {Tarat}}\ and\ \bibinfo {author} {\bibfnamefont {P.}~\bibnamefont
  {Majumdar}},\ }\href {\doibase 10.1140/epjb/e2015-50284-6} {\bibfield
  {journal} {\bibinfo  {journal} {Eur. Phys. J. B}\ }\textbf {\bibinfo {volume}
  {88}},\ \bibinfo {pages} {68} (\bibinfo {year} {2015})}\BibitemShut {NoStop}%
\bibitem [{\citenamefont {Karmakar}\ and\ \citenamefont
  {Majumdar}(2016{\natexlab{b}})}]{fflo}%
  \BibitemOpen
  \bibfield  {author} {\bibinfo {author} {\bibfnamefont {M.}~\bibnamefont
  {Karmakar}}\ and\ \bibinfo {author} {\bibfnamefont {P.}~\bibnamefont
  {Majumdar}},\ }\href {\doibase 10.1140/epjd/e2016-70250-2} {\bibfield
  {journal} {\bibinfo  {journal} {Eur. Phys. J. D}\ }\textbf {\bibinfo {volume}
  {70}},\ \bibinfo {pages} {220} (\bibinfo {year}
  {2016}{\natexlab{b}})}\BibitemShut {NoStop}%
\bibitem [{\citenamefont {Mukherjee}\ \emph {et~al.}(2016)\citenamefont
  {Mukherjee}, \citenamefont {Patel}, \citenamefont {Moreo},\ and\
  \citenamefont {Dagotto}}]{two-orb-1}%
  \BibitemOpen
  \bibfield  {author} {\bibinfo {author} {\bibfnamefont {A.}~\bibnamefont
  {Mukherjee}}, \bibinfo {author} {\bibfnamefont {N.~D.}\ \bibnamefont
  {Patel}}, \bibinfo {author} {\bibfnamefont {A.}~\bibnamefont {Moreo}}, \ and\
  \bibinfo {author} {\bibfnamefont {E.}~\bibnamefont {Dagotto}},\ }\href
  {\doibase 10.1103/PhysRevB.93.085144} {\bibfield  {journal} {\bibinfo
  {journal} {Phys. Rev. B}\ }\textbf {\bibinfo {volume} {93}},\ \bibinfo
  {pages} {085144} (\bibinfo {year} {2016})}\BibitemShut {NoStop}%
\bibitem [{\citenamefont {{Swain}}\ and\ \citenamefont
  {{Majumdar}}(2016)}]{2orb-2}%
  \BibitemOpen
  \bibfield  {author} {\bibinfo {author} {\bibfnamefont {N.}~\bibnamefont
  {{Swain}}}\ and\ \bibinfo {author} {\bibfnamefont {P.}~\bibnamefont
  {{Majumdar}}},\ }\href@noop {} {\bibfield  {journal} {\bibinfo  {journal}
  {arXiv:1610.00695}\ } (\bibinfo {year} {2016})}\BibitemShut {NoStop}%
\bibitem [{\citenamefont {Kumar}\ and\ \citenamefont {Majumdar}(2006)}]{kumar}%
  \BibitemOpen
  \bibfield  {author} {\bibinfo {author} {\bibfnamefont {S.}~\bibnamefont
  {Kumar}}\ and\ \bibinfo {author} {\bibfnamefont {P.}~\bibnamefont
  {Majumdar}},\ }\href@noop {} {\bibfield  {journal} {\bibinfo  {journal} {Eur.
  Phys. J. B}\ }\textbf {\bibinfo {volume} {50}},\ \bibinfo {pages} {571}
  (\bibinfo {year} {2006})}\BibitemShut {NoStop}%
\bibitem [{\citenamefont {Staudt}\ \emph {et~al.}(2000)\citenamefont {Staudt},
  \citenamefont {Dzierzawa},\ and\ \citenamefont {Muramatsu}}]{muramatsu-1}%
  \BibitemOpen
  \bibfield  {author} {\bibinfo {author} {\bibfnamefont {R.}~\bibnamefont
  {Staudt}}, \bibinfo {author} {\bibfnamefont {M.}~\bibnamefont {Dzierzawa}}, \
  and\ \bibinfo {author} {\bibfnamefont {A.}~\bibnamefont {Muramatsu}},\ }\href
  {\doibase 10.1007/s100510070120} {\bibfield  {journal} {\bibinfo  {journal}
  {Eur. Phys. J. B}\ }\textbf {\bibinfo {volume} {17}},\ \bibinfo {pages} {411}
  (\bibinfo {year} {2000})}\BibitemShut {NoStop}%
\bibitem [{\citenamefont {Paiva}\ \emph {et~al.}(2001)\citenamefont {Paiva},
  \citenamefont {Scalettar}, \citenamefont {Huscroft},\ and\ \citenamefont
  {McMahan}}]{scalettar-1}%
  \BibitemOpen
  \bibfield  {author} {\bibinfo {author} {\bibfnamefont {T.}~\bibnamefont
  {Paiva}}, \bibinfo {author} {\bibfnamefont {R.~T.}\ \bibnamefont
  {Scalettar}}, \bibinfo {author} {\bibfnamefont {C.}~\bibnamefont {Huscroft}},
  \ and\ \bibinfo {author} {\bibfnamefont {A.~K.}\ \bibnamefont {McMahan}},\
  }\href {\doibase 10.1103/PhysRevB.63.125116} {\bibfield  {journal} {\bibinfo
  {journal} {Phys. Rev. B}\ }\textbf {\bibinfo {volume} {63}},\ \bibinfo
  {pages} {125116} (\bibinfo {year} {2001})}\BibitemShut {NoStop}%
\bibitem [{\citenamefont {Mahan}(1990)}]{mahan}%
  \BibitemOpen
  \bibfield  {author} {\bibinfo {author} {\bibfnamefont {G.~D.}\ \bibnamefont
  {Mahan}},\ }\href@noop {} {\emph {\bibinfo {title} {{Quantum Many Particle
  Physics}}}}\ (\bibinfo  {publisher} {Plenum Press, New York},\ \bibinfo
  {year} {1990})\BibitemShut {NoStop}%
\bibitem [{\citenamefont {Kumar}\ and\ \citenamefont
  {Majumdar}(2005{\natexlab{b}})}]{conductivity}%
  \BibitemOpen
  \bibfield  {author} {\bibinfo {author} {\bibfnamefont {S.}~\bibnamefont
  {Kumar}}\ and\ \bibinfo {author} {\bibfnamefont {P.}~\bibnamefont
  {Majumdar}},\ }\href@noop {} {\bibfield  {journal} {\bibinfo  {journal} {Eur.
  Phys. J. B}\ }\textbf {\bibinfo {volume} {46}},\ \bibinfo {pages} {237}
  (\bibinfo {year} {2005}{\natexlab{b}})}\BibitemShut {NoStop}%
\end{thebibliography}%
\end{document}